\documentclass{aa}  
\usepackage{graphicx}
\usepackage{color}
\usepackage{hyperref}
\usepackage{float}
\usepackage{pbsi}
\usepackage{xcolor}
\usepackage{natbib}
\usepackage{epsf} 

\def \MJ{M$_{\mathrm{Jup}}$}

\def \ME{M$_{\oplus}$}

\def \RS{R$_{\odot}$}
\def \MS{M$\mathrm{_\odot}$}

\def \1s{$1\,\sigma$}

\def \t0{T$_0$}

\def \radvel{\texttt{RadVel}}

\begin{document}

\title{Characterizing planetary systems with SPIRou: questions about the magnetic cycle of 55 Cnc A and two new planets around B\thanks{Based on observations obtained at the Canada-France-Hawaii Telescope (CFHT) which is operated by the National Research Council (NRC) of Canada, the Institut National des Sciences de l'Univers of the Centre National de la Recherche Scientifique (CNRS) of France, and the University of Hawaii. The observations at the CFHT were performed with care and respect from the summit of Maunakea which is a significant cultural and historic site. } }

    \author{
C. Moutou\inst{1} \and
P. Petit\inst{1} \and
P. Charpentier\inst{1} \and 
P. Cristofari\inst{2} \and
C. Baruteau\inst{1} \and
P. Th\'ebault\inst{3} \and 
L. Arnold\inst{4}  \and 
E. Artigau\inst{5} \and
A. Carmona\inst{6,1} \and 
N.J. Cook\inst{5} \and
F. Debras\inst{1} \and
X. Delfosse\inst{6} \and
J.-F. Donati\inst{1} \and
L. Malo\inst{5} \and
M. Ould-Elhkim\inst{1} 
}

\institute{
\inst{1} Université de Toulouse, CNRS, IRAP, 14 avenue Belin, 31400 Toulouse, France, \email{claire.moutou@irap.omp.eu} \\
\inst{2} Laboratory for Astrophysics, Leiden Observatory, P.O. Box 9513, 2300 RA Leiden, The Netherlands \\
\inst{3} LIRA, Observatoire de Paris, Université PSL, 5 Place Jules Janssen, 92195 Meudon, France\\
\inst{4} Canada-France-Hawaii Telescope, CNRS, 96743 Kamuela, Hawaii, USA\\
\inst{5} Institut Trottier de Recherche sur les Exoplanètes, Université de Montréal, 1375 Ave Thérèse-Lavoie-Roux, Montréal, QC H2V 0B3, Canada\\
\inst{6} Université Grenoble Alpes, CNRS, IPAG, 38000 Grenoble, France\\
}

\date{June 2025}

\abstract{One of the first exoplanet hosts discovered thirty years ago, the star 55 Cnc has been constantly observed ever since. It is now known to host at least five planets with orbital periods ranging from 17 hours to 15 years. It is also one of the most extreme metal rich stars in the neighbourhood and it has a low-mass secondary star. In this article, we present data obtained at the Canada-France-Hawai'i Telescope with the SPIRou spectropolarimeter on both components of the 55 Cnc stellar system. We revisit the long-period radial-velocity signals of 55 Cnc A, with a focus on the role of the magnetic cycle, and propose the existence of a sixth planet candidate, whose period falls close to that of the magnetic cycle, or half of it. The other massive outer planet has a revised period of 13.15 years and a minimum mass of 3.8 \MJ. Although some uncertainty remains on these outer planets, the characterization of the four inner planets is very robust through the combination of many different data sets, and all signals are consistent in the nIR and optical domains. In addition, the magnetic topology of the solar-type primary component of the system is observed by SPIRou at the minimum of its activity cycle, characterized by an amplitude ten times smaller than observed during its maximum in 2017. For the low-mass component 55 Cnc B, we report the discovery of two exoplanets in the system, with a period of $6.799\pm0.0014$ and $33.75\pm 0.04$ days and a minimum mass of 3.5$\pm$0.8 and $5.3\pm 1.4$ \ME, respectively. The secondary magnetic field is very weak and the current data set does not allow its precise characterization, setting an upper limit of 10~G. The system 55 Cnc stands out as the sixth binary system with planetary systems around both components, and the first one with non equal-mass stellar components.}

\titlerunning{The system of 55 Cnc }
\authorrunning{Moutou et al}
\maketitle

\section{Introduction}
The nearby Sun-like star 55 Cnc (also named $\rho^1$ Cnc, HD 75732, HIP 43587, HR 3522) hosts five known exoplanets. Its planet b, a giant planet in a 14.65~d orbital period, was the fourth exoplanet ever discovered, from Lick Observatory radial velocity (RV) measurements \citep{butler1997}. A few years later, new measurements indicated the presence of other giant planets in the system, notably in the 5110~d and 44~d periods \citep{marcy2002}. The innermost Neptune-like planet was first found in a 2.8~d period \citep{McArthur2004,fischer2008}, which turned out to be the 1~d alias of an even closer and less massive planet at 0.7365~d \citep{dawson2010}. This inner planet was then confirmed for its extremely short period and found to be transiting \citep{winn2011}. \citet{fischer2008} further identified a 0.15 M$_{Jup}$ planet signal at 260~d. These five planets are thus sorted e-b-c-f-d in order of their orbital periods. The gap between planets f and d is of about 4 au. From Doppler noise estimates in multiple instrument datasets, \citet{baluev2015} finds indications for an activity cycle of 4599~d duration in addition to a planet d with a 4825~d period. Two recent RV analyzes, including new RV data, revisit the longest-period signals. \citet{bourrier2018} uses the long-period periodicity of stellar chromospheric indices to fix a 3822~d magnetic cycle (using a Keplerian model) and infer the residual RV signal to planet d at 5574$\pm$90~d. Then, \citet{rosenthal2021} identified two long-period signals at 1966$\pm$23~d and 6110$\pm$200~d and assigned the first one to a systematic instrumental effect and the second one to the magnetic cycle signature, in addition to the 5218$\pm$200~d planet d. Finally, \citet{laliotis2023} also provides an update of the data, showing a variation of the $S$ index with a period of 3801$\pm$130~d interpreted as the activity cycle and no update on the long-period planet d. The identification of the longest-period signals is thus controverted, both because of the complexity of mixing RV data from multiple instruments and the ill-constrained impact of the magnetic cycle. The period of the $\sim$15y period planet is also more difficult to constrain in this context.
Concerning radial velocities, literature data are thus abundant for this bright northern star, observed by many different instruments. A comprehensive compilation has been provided by \citet{bourrier2018} and additional HIRES and APF RV data are published in \citet{rosenthal2021}.

A less massive star is also part of the stellar system 55 Cnc, a mid-M dwarf located at a projected separation of 1065 au, far enough from the primary so that the latter's perturbations should not inhibit planet formation around this companion \citep[see review on planet formation in binaries in][]{theb2015}. The secondary star has been less well studied than its brighter, solar-like counterpart. 

In this article, we study the stellar and planetary system as a whole, using recent SPIRou data on both stars and collecting data from the literature. In Section 2, we review the fundamental parameters of both stars. In Section 3, we present SPIRou observations obtained in the last years. In Sections 4 and 5, we present our analysis of the magnetic properties of both stars and of both RV time series. Finally, we discuss some system's characteristics in Section 6 and draw our conclusions in Section 7.

\begin{table}
\caption{Adopted parameters of target stars\label{table:parameters}}
\begin{tabular}{lccccc}
\hline\hline
Name  & Mass & Radius&  SpT & P$_{\rm rot}$ \\
      & (\MS) & (\RS) & & (d)  \\\hline
55 Cnc A  &1.015$\pm$0.051 & 0.980$\pm$0.016& G8V   & 39$\pm$1\\
55 Cnc B & 0.26$\pm$0.02 & 0.274$\pm$0.066 & M4.5V &  92$\pm$5  \\\hline
\end{tabular}
\end{table}x

\begin{figure}[ht!]
    \centering
\includegraphics[width=\hsize]{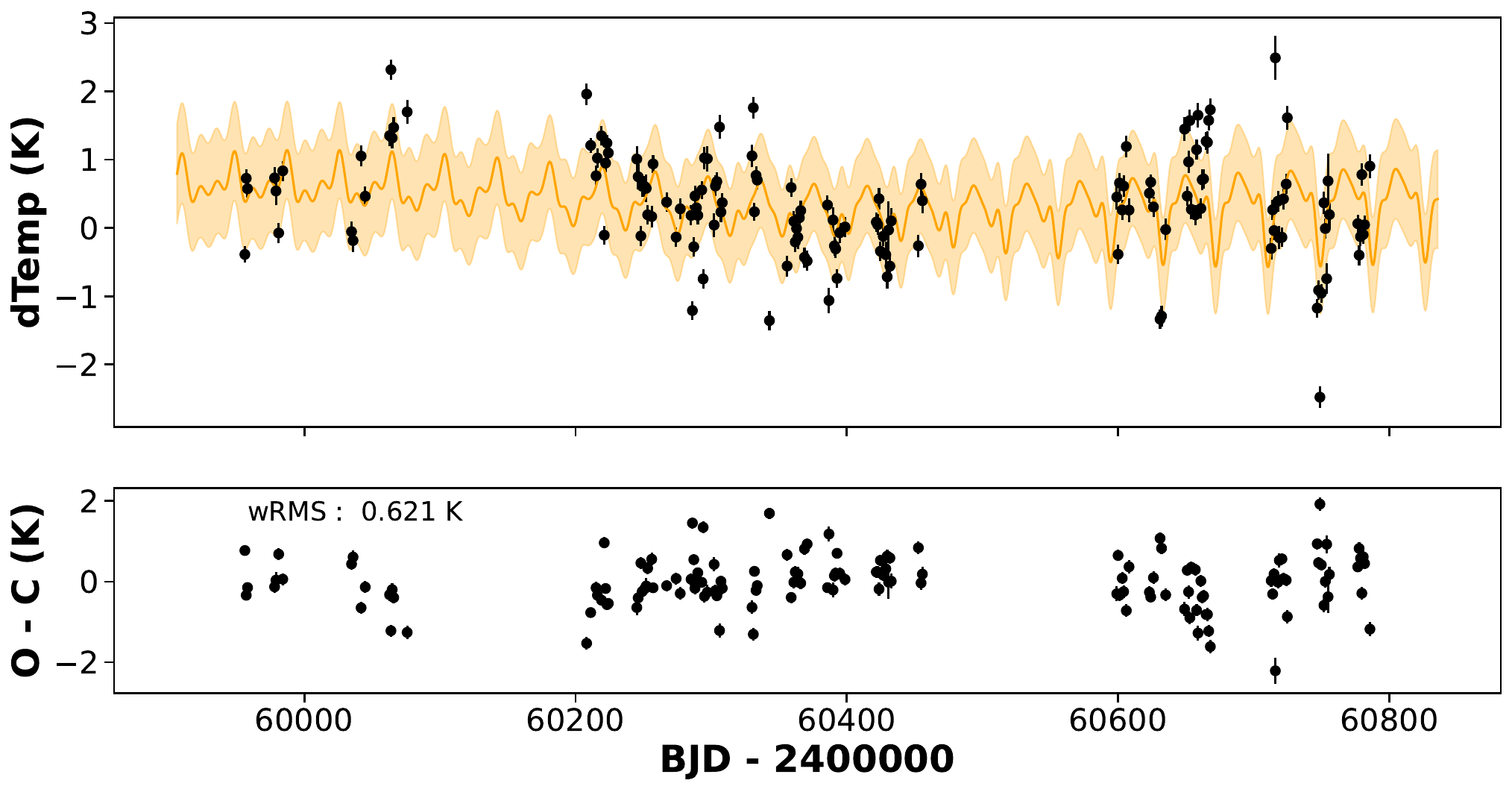}
\includegraphics[width=\hsize]{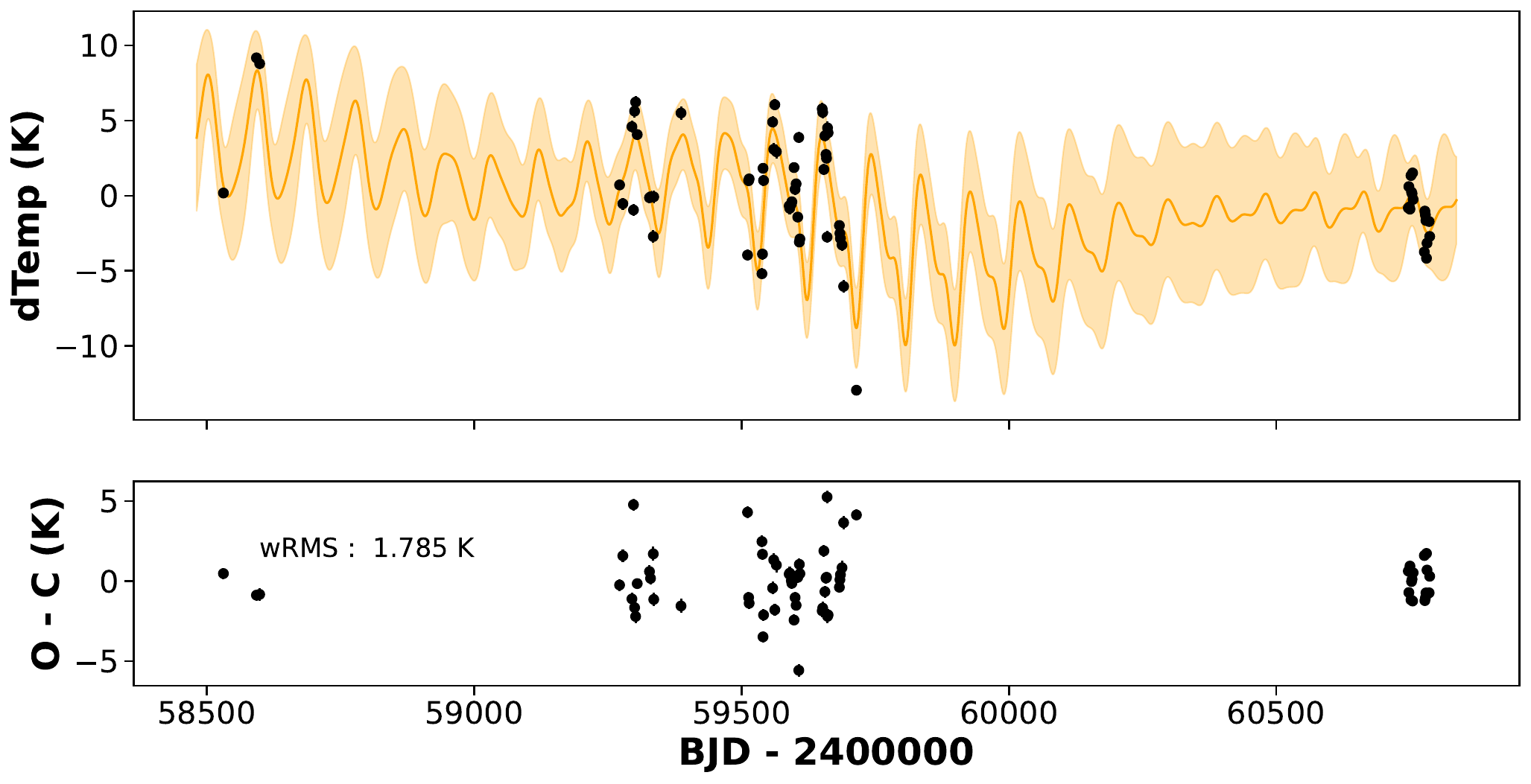}
\caption{Quasi-periodic GPR model of the differential temperature time series of 55 Cnc A (top) and B (bottom). }\label{fig:dt}
\end{figure}

\begin{table}
	\centering
	\caption{GPR prior and posterior parameters for the differential temperature variations }
	\begin{tabular}{lcc} 
		\hline
		\hline
		Parameter       & Prior                        & Posterior  \\ \hline &55 Cnc A &\\\hline
		a (K)           & $\mathcal{MLU}(0.1,10)$       &  0.52$_{-0.18}^{+0.49}$\\
		$\lambda_e$ (d) & Fixed (300)                  & 300  \\
		$P_{\rm rot}$ (d)   & $\mathcal{N}(38.8,2) $   &  38.7$_{-0.5}^{+0.8}$\\
		$\lambda_p$     &  $\mathcal{LU}(0.1,5)$  &  0.74$_{-0.40}^{+2.47}$ \\
		Jitter (K)       &  $\mathcal{MLU}(0.,10)$   &  0.66$\pm0.05$  \\ \hline &55 Cnc B  &\\\hline
		a (K)           & $\mathcal{MLU}(0.1,30)$     & $5.0_{-1.3}^{+2.1}$ \\
		$\lambda_e$ (d) & Fixed (300)                  & 300  \\
		$P_{\rm rot}$ (d)   & $\mathcal{U}(3, 140)$    & $92_{-5}^{+4}$    \\
		$\lambda_p$     & $\mathcal{LU}(0.1,5)$  & $1.11_{-0.39}^{+0.72}$ \\
		Jitter (K)      & $\mathcal{MLU}(0.1,30)$  & $2.1\pm0.26$  \\ \hline
	\end{tabular}
	\label{tab:GPR_dt_55CNCAB}
\end{table}

\section{Stars}
\subsection{55 Cnc A}
The 55 Cnc distance has been updated by $Gaia$ measurements with a distance of 12.590$\pm$ 0.012\,pc \citep{GaiaDR3}.
The host star 55 Cnc is known to be a super-metal-rich star; as a $Gaia$ benchmark star, it has been scrutinized and its [Fe/H] value has been recently updated to 0.32$\pm$0.02 \citep{soubiran2024}, in agreement with the literature values ranging from 0.25 to 0.45. Apart from its metallicity, the stellar fundamental properties are close to those of the Sun with a stellar density of 1.079$\pm$0.005 times the solar density and a mass of 1.015$\pm$0.051\,\MS\ \citep{crida2018}, constrained by the precise stellar density derived from transits of planet e, luminosity, and interferometric radius measurements. Stellar evolution models predicted two separate solutions at 13.2\,Gy and 31\,Myr \citep{ligi2016} with a preference for the young scenario corresponding to the interferometrically-derived stellar radius, while the old scenario is clearly more in line with the observed very quiet chromospheric star (log$R'_{\rm HK}$ of -5.0 and brightness variations of the order of 1\,mmag, as in \citet{marcy2002,bourrier2018}). In particular, a rotation period larger than 30 days (see below) is not expected for a 30 Myr solar-type star \citep[e.g.,][]{barnes2007,gallet2013}. It is possible that stellar models are less accurate for such high stellar metallicity and fail at predicting the observed mass and radius without ambiguity. We opt for the configuration where the age is greater than 10 Gy.

A mean spectrum of 55 Cnc A (called template hereafter) is obtained from all available spectra observed with SPIRou after instrumental and telluric corrections are applied (see section 3.1). For 55 Cnc A, more than 1000 individual spectra were summed up.
Using this stellar template spectrum, we then used $Zeeturbo$ to derive the fundamental parameters as well as the small-scale magnetic field of the primary star, as described in \citet{cristofari2023}. We derive an effective temperature of 5198$\pm$31\,K, log$g$ of 4.30$\pm$0.05, a metallicity [M/H] of 0.39$\pm$0.10\,dex in agreement with \citet{soubiran2024}. The model requires a line broadening that is obtained by a v$sin$i of 1.06$\pm$0.23\,km/s with a macroturbulence velocity of 2.86\,km/s, and a small-scale magnetic field of 250$\pm$20\,G. More precisely, we find out that the filling factor with a null magnetic field represents 87\% while the bin with 2\,kG magnetic field represents 13\% of the surface. Due to the low magnetic field, no component of higher field strength were included in the fit.

The stellar rotation period is derived from chromospheric or photometric variations to be in the range of 34 to 43\,d \citep{bourrier2018,laliotis2023}. Such a range may be indicative of a certain level of differential rotation, expected in solar-like stars. A rotation period of 39\,d was measured by Zeeman Doppler Imaging (ZDI) in a spectropolarimetric campaign conducted in 2017 \citep{folsom2020} and in H$\alpha$ equivalent-width time series \citep{bourrier2018}, and this is the value we adopt here. The ZDI analysis also places an upper limit on differential rotation at a level of 0.065 rad.day$^{-1}$ \citep{folsom2020}.

There is a necessity to better understand what is due to the magnetic activity in the long-term RV modulation. Spectropolarimetry may help. The cycle of a Sun-like star is visible in its circular polarization map along the years, as can be observed with spectropolarimeters (e.g. \citealt{borosaikia2016, mengel2016}). There is a previous magnetic map available for 55 Cnc, obtained from TBL/NARVAL data in 2017 and analyzed by \citet{folsom2020}. The reconstructed large-scale magnetic field of 55 Cnc was then found to be mostly poloidal, with 80 (20)\% of the magnetic energy in the dipole (respectively, the quadrupole) and an average strength of 3.4 G. The stellar wind modelling gave a value of 55 Cnc mass loss similar to the Sun's with $\sim 2 \times 10^{-14}$ \MS yr$^{-1}$.

\subsection{55 Cnc B}
The stellar companion 55 Cnc B is an M4V star at 1065\,au South-East from the most massive component (84" visual separation) \citep{eggenberger2004}. It has a $V$ magnitude of 13.15 and a $H$ magnitude of 7.93. An effective temperature of 3187$\pm$157~K and a mass of 0.265$\pm$0.02\,\MS\ are reported in the revised TESS Input Catalog by \citet{stassun2019}.
A high metallicity of [M/H]$=$0.42 is reported by \citet{Houdebine2016} and an effective temperature of 3226\,K in \citet{Houdebine2019}. However, a wide range of temperatures and metallicities is present in the literature for that star, spanning from 3000 to 3450\,K and from -0.2 to +0.5 dex, respectively\footnote{Values retrieved from the SIMBAD database, available at https://simbad.cds.unistra.fr/simbad}. A significant discrepancy of the metallicities for stars in this system is reported by \citet{marfil2021}, with a negative metallicity reported for 55 Cnc B.
 
The SPIRou template spectrum was analyzed using $Zeeturbo$ as in \citet{cristofari2023}, providing us with a global fit of fundamental parameters and the small-scale magnetic field that may impact the magnetically sensitive lines of the star. This template was obtained from the first 52 SPIRou spectra (see section 3.1). This analysis gives an effective temperature 3286$\pm$31\,K, log$g$ of 4.86$\pm$0.05 and [M/H] of 0.16$\pm$0.10 with a  vsin$i$ of 1.29$\pm$0.31\,km/s. Here again, a macroturbulence velocity was fixed to 2.9\,km/s for this fit. Figure \ref{fig:corner55cncB} shows the posterior distribution of these parameters.
The metallicity derived for the two components agrees within 2$\sigma$ and does not confirm the extreme discrepancy reported by \citet{marfil2021}.

For this star, we can only set an upper limit on the unsigned, small-scale magnetic field at $<90$\,Gauss (3-$\sigma$ confidence), meaning that less than 1\% of its surface has a field larger than 2\,kG (parameter a$_2$ in the corner plot). This is the signature of a quiet M star as compared to the 44 M star sample of the SPIRou Legacy Survey \citep{cristofari2023a}, confirming the "old system" option.

Using the absolute magnitude in the $K$ band and the calibrated mass-luminosity relations of \citet{mann2019}, one can derive the stellar mass of 55 Cnc B as 0.26$\pm$0.02\,\MS. The stellar radius is obtained from the estimated effective temperature, absolute magnitudes, and bolometric corrections \citep{cifuentes2020}, and we find 0.27$\pm$0.07\,\RS. There is then a 2$\sigma$ difference on log$g$ value, with log$g$ derived from these mass and radius values of 4.98 instead of 4.86 as previously fitted. 

Rotation modulation has been searched for in Kepler data, with a non-detection and an estimate of a period longer than 500\,d by \citet{newton2016}. In \citet{Houdebine2016}, a minimum value of P$_{rot}$/sin$i$ of 6.11$\pm$3\,d is derived from the rotational velocity, which is known to be poorly adapted for slow rotators due to insufficient spectral resolution. Such discrepancy may be explained by the extremely low surface activity level and subsequent non-detection of a rotational modulation.

\begin{table}
	\centering
\caption{Summary of SPIRou observations}.\label{table:spiroudata}
\begin{tabular}{lcc}
\hline\hline
Name & 55 Cnc A & 55 Cnc B \\\hline
N&              141 & 69\\
<SNR>&        510 & 180\\
n$_\gamma$ ($Y$) m/s  & 1.98  & 14.8\\
n$_\gamma$ ($J$) m/s   & 2.01  & 12.9\\
n$_\gamma$ ($H$) m/s   & 1.49  & 7.7\\
n$_\gamma$ ($K$) m/s   & 4.72  & 7.9 \\
n$_\gamma$ ($YJHK$) m/s & 0.92 & 1.8 \\
$\sigma_{\rm RV}$ m/s  & 50.98 & 5.9\\\hline
\end{tabular}
  \tablefoot{N is the number of visits, <SNR> is the mean SNR per sequence in the $H$ band, n$_\gamma$ is the RV noise per band separately and combined, and $\sigma_{\rm RV}$ is the dispersion}
\end{table}

\section{Data}

\subsection{SPIRou}
New data on the stellar system of 55 Cnc have been collected with the infrared spectropolarimeter SPIRou mounted on the Canada-France-Hawaii 3.6-m telescope atop Maunakea, Hawai'i. 
SPIRou observes in one shot the spectral domain from 0.98 to 2.35 $\mu$m at a resolution power of 70k. With its stabilized cryogenic spectrograph and simultaneous wavelength calibration, it is used for precision RV measurements, while the Cassegrain polarimetric unit allows recording the large-scale magnetic field of the star through circularly-polarized light at the same time \citep{donati2020}. An observation consists of a series of four consecutive exposures during which the Fresnel rhombs are rotated in such a way that the circularly polarized spectrum and the Stokes V profile can be derived \citep{donati1997}.

SPIRou spectra and associated calibrations have been reduced using the APERO pipeline \citep{cook2022} and RVs were extracted with the line-by-line (LBL) method adapted to SPIRou by \citet{artigau2022}. 
Whenever necessary, the systematics in the data was then suppressed by the Wapiti software that uses the principal component analysis in the LBL framework \citep{ouldelhkim2023}.
Moreover, differential effective temperatures from LBL derived products were measured following the method described by \citet{artigau2024}.

The more massive component of the system, 55 Cnc A, was observed as a sun-like calibration star by the observatory (program ID Q57 from semesters 22B to 25A). The observations used in the current study are done in spectropolarimetric mode with 151 sequences collected from 11 January 2023 to 20 April 2025. Each observation consists of four consecutive exposures of 61.29 s. The SNR per pixel in the $H$ band ranges from 190 to 310 per exposure and ten exposures have significantly lower SNR values and are discarded in the analysis.
Note that other data of 55 Cnc A have been obtained with SPIRou for atmospheric studies of planet e; they were not included in the study here because they either do not use the simultaneous wavelength calibration, or do not use the circular polarization observing mode, or both. However, they do contribute to having a better stellar template for 55 Cnc A by co-adding 1032 individual spectra, which in turn improves the telluric correction of each spectrum and, ultimately, RV precision.

The secondary M star was observed through the "small sisters" program (PI C. Moutou, program IDs 19AF02, 21AF02, 21BF17, 22AF09, 25AD08), a program aiming at searching for planets around M stars that are companions to more massive known planet hosts. Seventy polarimetric sequences were collected from 16 February 2019 to 21 April 2025.

Table \ref{table:parameters} lists the final stellar parameters adopted for this study.

\subsection{Archival data}
In addition to the SPIRou data, various archival, optical spectra were collected for the chromospheric activity indicator (see Sect. 4.1), longitudinal magnetic field measurements (see Sect. 4.2) and RV analysis (see Sects. 5.1 and 5.2), showing the interest of long-term open-access data sets. The origin of these data is detailed out in the relevant sections.

\subsection{Differential temperature}
When projected on a library of stellar templates, SPIRou line-by-line information can be used to retrieve a disk-averaged differential temperature (dTemp) indicator, which is a proxy for the star's surface activity. The method is explained in \citet{artigau2024} where it is valid in the effective temperature range of 3000 to 6000\,K. Both components of 55 Cnc are at the edges of the validity domain, but looking at these data can nevertheless be instructive. 

We modelled the tiny variations of 55 Cnc A dTemp time series  with a quasi-periodic Gaussian Process Regression (QP-GPR) seen with the 5500 and 6000\,K templates (see Figure \ref{fig:dt}). We used a kernel whose covariance function is described as: 

\begin{equation}
    Q_{i,j} = a \times exp \left( - \frac{\sin^2 \left( {\frac{\pi \Delta t_{i,j}}{P_{\rm rot}}}\right)}{2 \lambda_p^2} - \frac{\Delta  t_{i,j}^2}{2 \lambda_e^2}   \right) + Jitter\label{eq:pyaneti2}
\end{equation}

with $P_{\rm rot}$ the rotation period, $\lambda_e$ the evolution timescale and $\lambda_p$ the smoothing factor. In the following, $\lambda_e$ is set to 300\,days. Priors for the amplitude and jitter values are set as modified log-uniform functions $\mathcal{MLU}$ with initial limits defined from the dispersion of the data. The prior of $\lambda_p$ is a logarithmic uniform function $\mathcal{LU}$. These choices are based on \citet{Camacho2023}.
Table \ref{tab:GPR_dt_55CNCAB} lists the prior and posterior parameters used in the QP-GPR fit. Figure \ref{fig:app1_dtA} shows the corner plot. The rotation period of 55 Cnc A is found to be 38.7$\pm 0.8$\, days.

Star 55 Cnc B shows a slightly more contrasted dTemp modulation when the 3000\,K or the 3500\,K templates are used. 
Figure \ref{fig:dt} shows a QP-GPR of 55 Cnc B dTemp time series with the 3500\,K template, which best fit gives an amplitude of $5\pm2$\,K and a period of $92\pm5$ days (see corner plot in Fig \ref{fig:app1_dtB}). We used a uniform prior from 3 to 140 days, since no previous estimate of the period was available in the literature.

\begin{figure}
    \centering
\includegraphics[width=1.1\hsize]{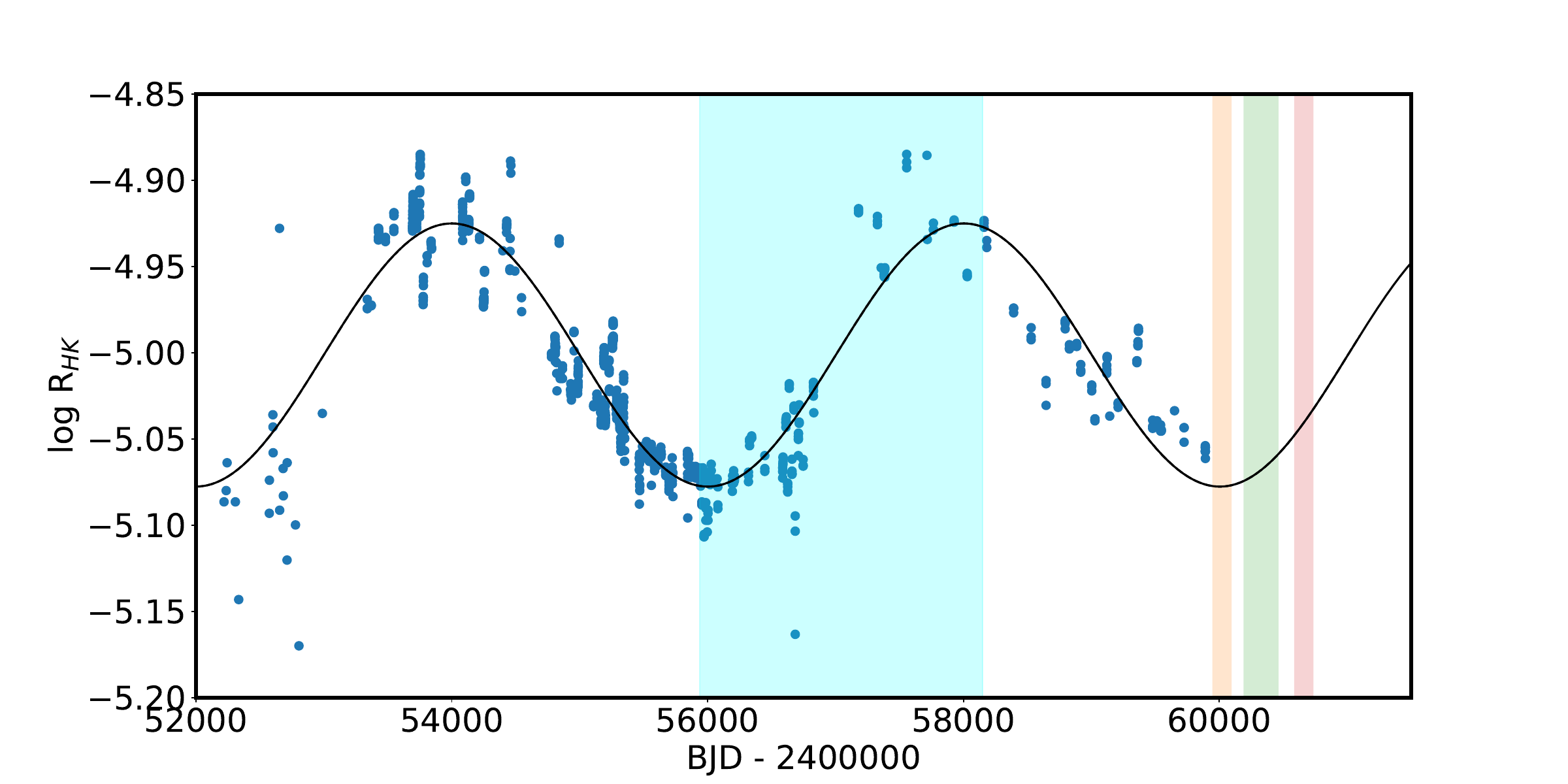}
\includegraphics[width=1.1\hsize]{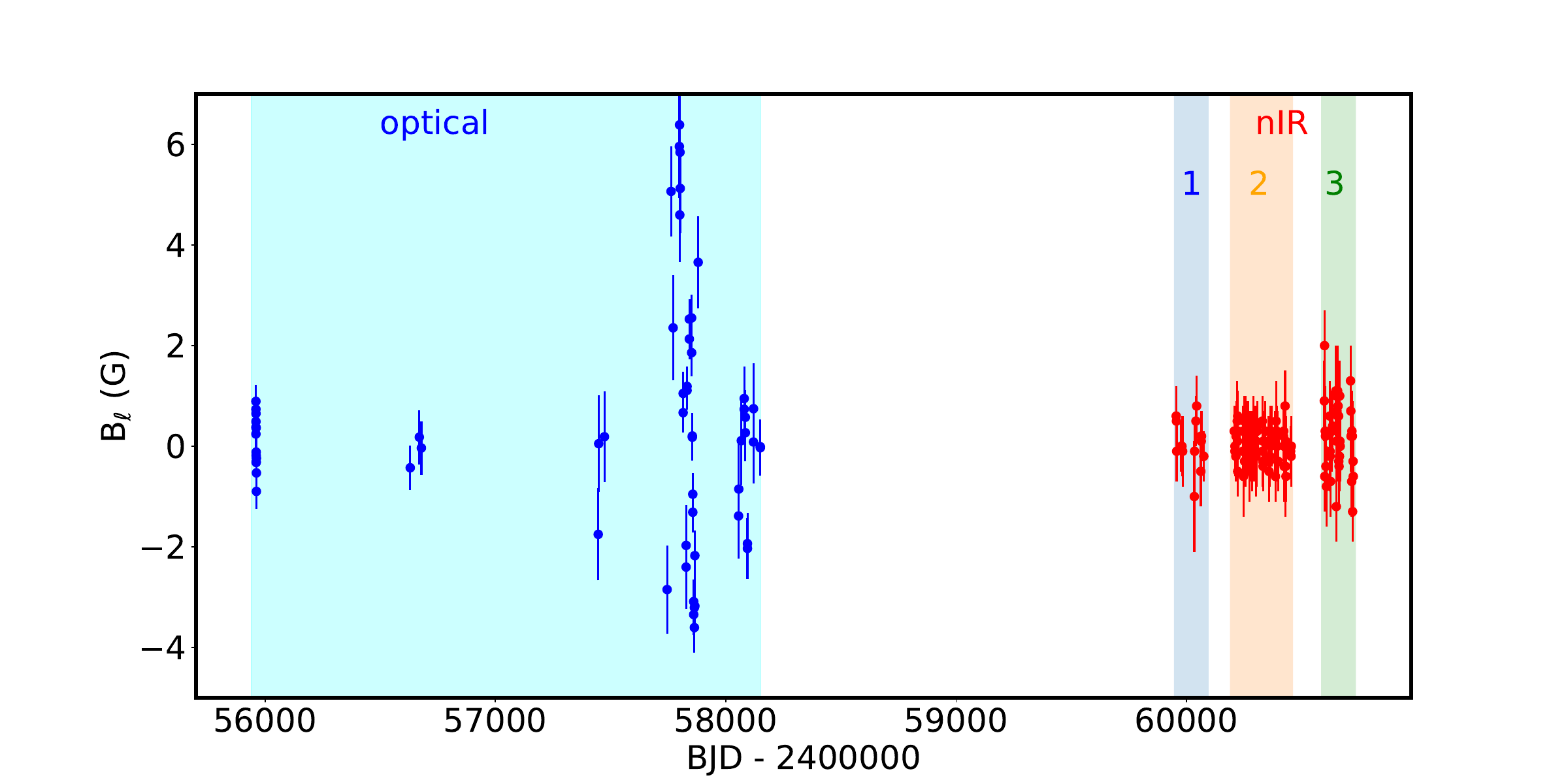}
\caption{Top: time variations of the  log$R'_{\rm HK}$ as measured by optical instruments for 55 Cnc A \citep{bourrier2018,isaacson2024}. The blue (resp. red) vertical lines show the range of optical (resp. nIR) spectropolarimetric campaigns. The best-fit sinusoidal model is overplotted, with a 4003d period. Bottom: longitudinal field measurements of 55 Cnc A as a function of the Julian date. Blue symbols show optical measurements, while red symbols display nIR observations. Shaded areas show the different subsets used for ZDI analysis, especially for SPIRou where the three season were treated separately (see text). One SPIRou measurement with an error larger than 3\,G was removed. Note the different time scales.}\label{fig:cycle_55cncA}
\end{figure}

\section{Magnetic field analyzes}

\subsection{The magnetic cycle of 55 Cnc A}

Figure \ref{fig:cycle_55cncA} shows when the spectropolarimetric data have been collected with respect to the chromospheric cycle observed in the Ca H and K lines with optical instruments (mainly HIRES, with contributions of HARPS and HARPS-N) \citep{bourrier2018,isaacson2024}. The chromospheric cycle can be fitted using a sinusoidal model with a period of 4003\,d, which is 2.4$\sigma$ longer than the 3822$\pm77$ d period used in \cite{bourrier2018} when only one period of the cycle was available. 

\subsection{Magnetic topology of 55 Cnc A seen by SPIRou}

SPIRou was setup in the polarimetric configuration when observing 55 Cnc A. A circularly polarized (Stokes V) spectrum was recorded and reduced for every visit, using the APERO reduction pipeline, leading to a total of 131 polarimetric observations. In addition to the near-infrared material, another 62 optical observations were obtained from the PolarBase archive \citep{2014PASP..126..469P}, with observations spread from 2012 to 2018 obtained by ESPaDOnS at CFHT, or its twin NARVAL at Telescope Bernard Lyot (France).

To increase the signal-to-noise ratio, all spectra were processed with the Least-Squares-Deconvolution method (LSD hereafter, \citealt{1997MNRAS.291..658D,2010A&A...524A...5K}), using an open source implementation of the algorithm\footnote{https://github.com/folsomcp/LSDpy}. The line list used for cross-correlation was extracted from the VALD database \citep{2015PhyS...90e4005R} using atmospheric parameters close to those of 55 Cnc A. We applied the $\chi^2$ criterion of \cite{1997MNRAS.291..658D} to look for the detection of polarized signatures in the line profile, which led to multiple detections of Zeeman signatures in the optical data set, while none of the 131 SPIRou visits led to a definite detection. However, we note that the latest SPIRou run (October 2024 to February 2025) produced statistics closer to the detection threshold.  

We computed longitudinal magnetic field ($B_l$) measurements \citep{1979A&A....74....1R}, using the specPolFlow dedicated package\footnote{https://github.com/folsomcp/specpolFlow} (Figure \ref{fig:cycle_55cncA}, bottom panel). A fair fraction of optical data display $B_l$ absolute values below 1~G, with a much larger dispersion observed between December 2016 and December 2017, leading to a number of $B_l$ absolute values above 2~G, and up to 6~G. This period of increased magnetism includes the data presented in the study of \cite{folsom2020}. It is also very close in time to the $R'_{\rm HK}$ maximum reported by \cite{isaacson2024}, confirming that Folsom's data were taken at activity maximum (Figure \ref{fig:cycle_55cncA}, top). In contrast, measurements from 2012, all below 1~G in absolute value, were representative of the $R'_{\rm HK}$ minimum. The SPIRou measurements started close to the following $R'_{\rm HK}$ minimum, while the last SPIRou campaign (Oct. 2024 to Feb. 2025) took place during the expected activity increase towards a future maximum. In agreement with this broad picture, early SPIRou measurements are consistent with archival measurements from the previous 2012 minimum. $B_l$ measurements from the latest SPIRou run show increased variability, while the error bars remain at their standard level (of around 0.6~G).      

We applied the Generalized Lomb-Scargle algorithm (GLS, \citet{Zechmeister2009}) to run a period search on the $B_l$ estimates. From the optical data, we obtain a period of 38.6~d, in good agreement with the 39~d period used by \cite{folsom2020}. Using SPIRou observations, the rotation period does not show up, whether we use the whole time series or subsets. Instead, a peak at around 22 days is recovered, possibly because of the quadrupolar component of the surface field.

We used the Zeeman-Doppler Imaging (ZDI) method \citep{1989A&A...225..456S,2006MNRAS.370..629D} on SPIRou data to reconstruct the large-scale magnetic geometry of 55~Cnc~A. We employ here the python implementation of \cite{2018MNRAS.474.4956F,2018MNRAS.481.5286F}. We first checked if, by making use of the full amount of information available in the Stokes V line profiles, the ZDI modelling was able to detect the rotational modulation of the magnetic signal, since relying on the first moment of Stokes V profiles ($B_l$ estimates, see above) proved to be inefficient. To do so, we split the full SPIRou data set into three subsets: 15 observations from Jan. 2023 to May 2023 as subset \#1, 76 observations from Sep. 2023 to May 2024 as subset \#2, and 40 observations from Oct. 2024 to Feb. 2025 as subset \#3. Separating in subsets allows characterizing the evolution of the magnetic maps over the years while minimizing the $\chi^2$. We checked, indeed, that modelling the whole time series in a single ZDI map did increase the $\chi^2$. The second season with more observations corresponds to the cycle minimum; a test was carried over to see whether observations in subset \#2 could be split into two even parts. The test showed that the two resulting maps were similar, and noisier. It is thus optimal to model one map per observing season.
We used an approach similar to \cite{2008MNRAS.388...80P,2022A&A...666A..20P} where ZDI models are computed for a range of rotation periods, looking for the period value that minimizes the $\chi^2$ of the model. The best model was found for a period of $37.9 \pm 1.1$, $39.8 \pm 0.7$, and $36.3 \pm 0.5$~d for subsets \#1, 2, and 3, respectively. Since these values are reasonably close to the expected rotation period, we conclude that a rotationally-modulated polarized signal can be recovered from the SPIRou observations, although it is dominated by the noise when individual visits are considered.

\begin{figure*}
\centering
\includegraphics[width=6cm]{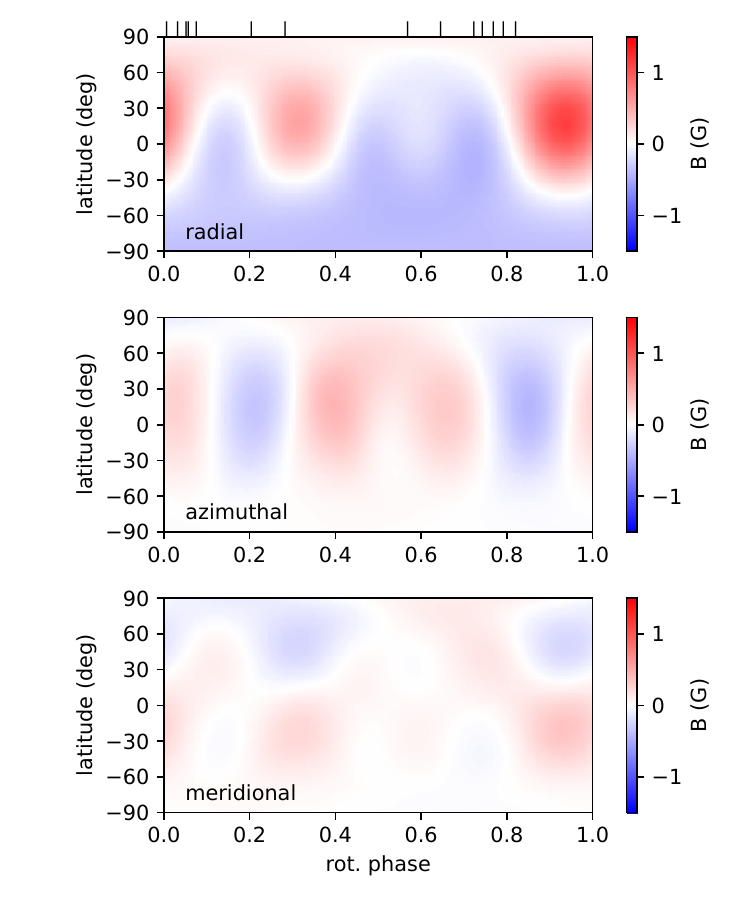}
\includegraphics[width=6cm]{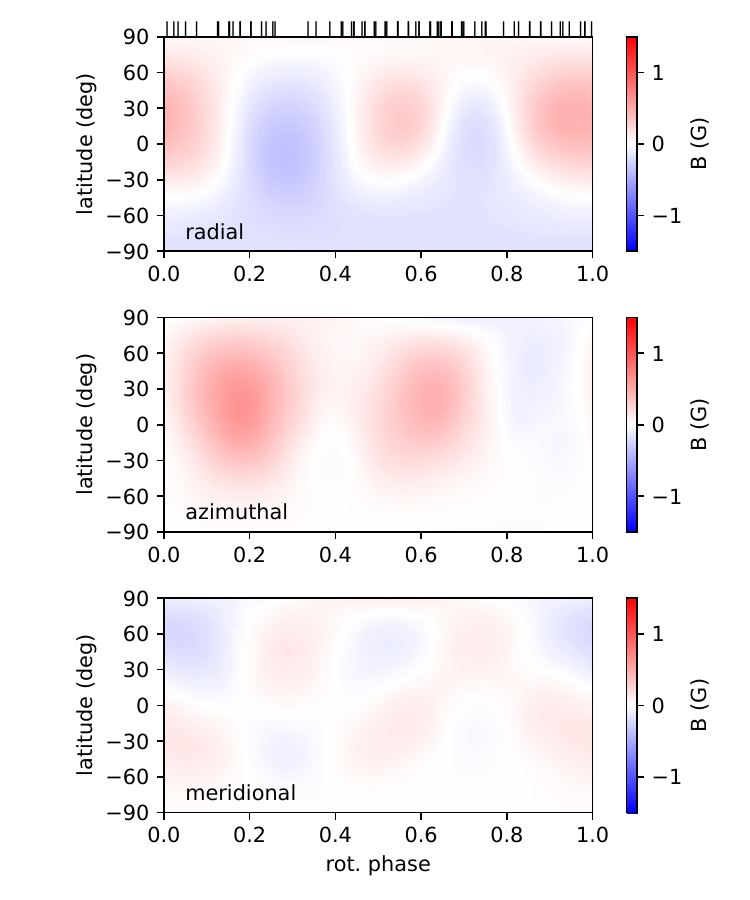}
\includegraphics[width=6cm]{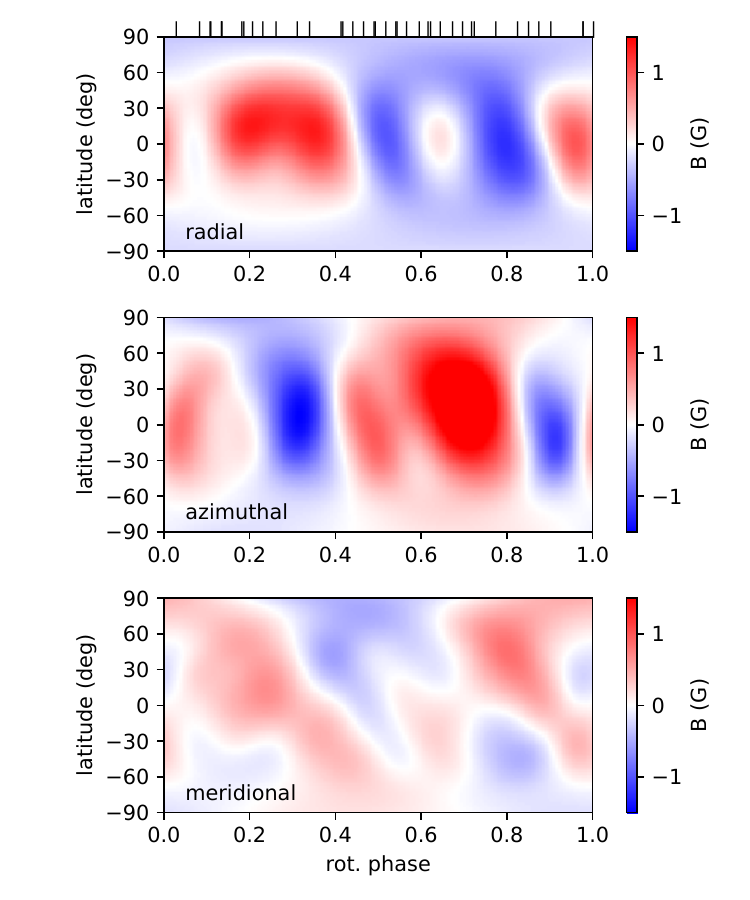}
\caption{Magnetic geometry of 55 Cnc A. From left to right, we show ZDI models from observations of Jan. to May 2023 (subset \#1), Sep. 2023 to May 2024 (subset \#2), Oct. 2024 to Feb. 2025 (subset \#3). The stellar surface is shown in equatorial projection, with phases of observation indicated as vertical ticks above the top panel. The vector magnetic field is projected onto a spherical frame, with the radial, azimuthal, and meridional components displayed from top to bottom. The field strength is expressed in Gauss. 
}
\label{fig:maps}
\end{figure*}

Following this test, we proceeded with the ZDI inversion, analysing each subset separately to account for the natural variability of the magnetic geometry. We set our input parameters to the values adopted by \cite{folsom2020}, with $P_{\rm rot} = 39$~d and $v \sin i = 1.2$~km~s$^{-1}$. Despite an inclination angle likely close to 90$^\circ$, we adopted a slightly lower value of 70$^\circ$ to limit the mirroring effect reported by Folsom. By doing so, we followed the strategy proposed by \cite{2023MNRAS.525..455D} for AU~Mic. The resulting maps are shown in Figure \ref{fig:maps}. For the subsets \#1 and \#2, we obtain an averaged unsigned field of 0.3~G, which is nearly ten times weaker than the value obtained by Folsom near activity maximum. For subset \#3, the average field increases to about 0.9~G, confirming that 55~Cnc~A left its activity minimum before the last SPIRou run. For all subsets, the field geometry is dominated by the poloidal component (storing between 70\% and 98\% of the magnetic energy). For subsets \#1 and 3, the main field component was a dipole, while the quadrupole dominated the geometry in subset \#2. At all epochs, the field geometry was mostly non-axisymmetric, with 9\% to 37\% of the magnetic energy in spherical harmonics modes with $\ell = 0$. We note, however, that the large inclination angle tends to hide any dipole aligned on the spin axis, since its polar field would be mostly perpendicular to the line of sight (while Stokes V is sensitive to the line-of-sight projection of the magnetic field), and the situation would be made worse by the cancelling polarities in the two hemispheres.  

\subsection{Magnetic topology of 55 Cnc B}

\begin{figure}[ht!]
    \centering
\includegraphics[width=\hsize]{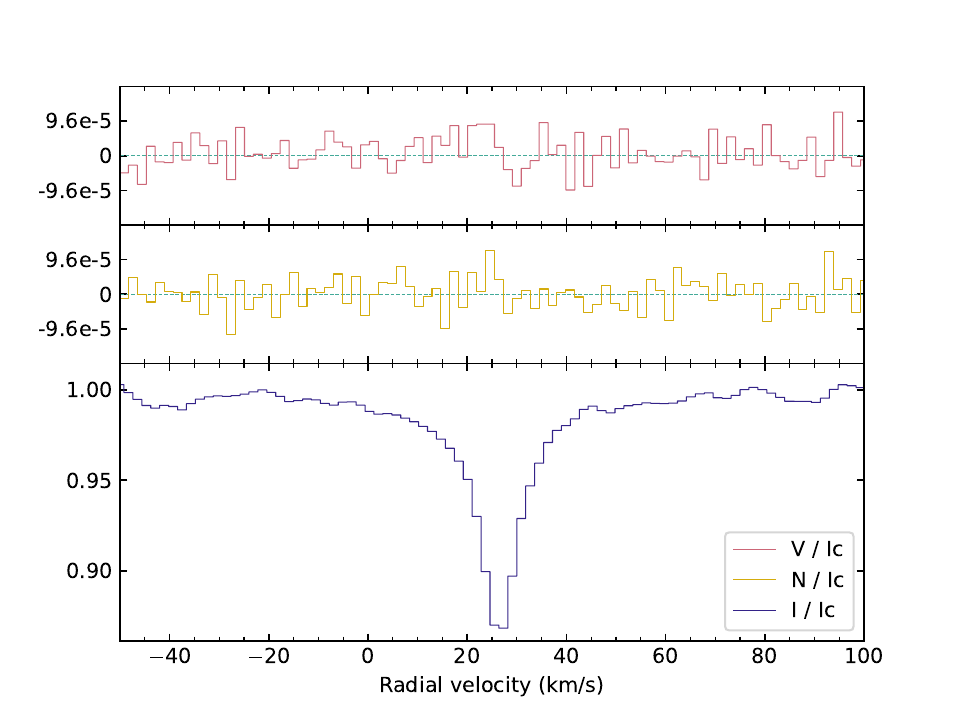}
\includegraphics[width=\hsize]{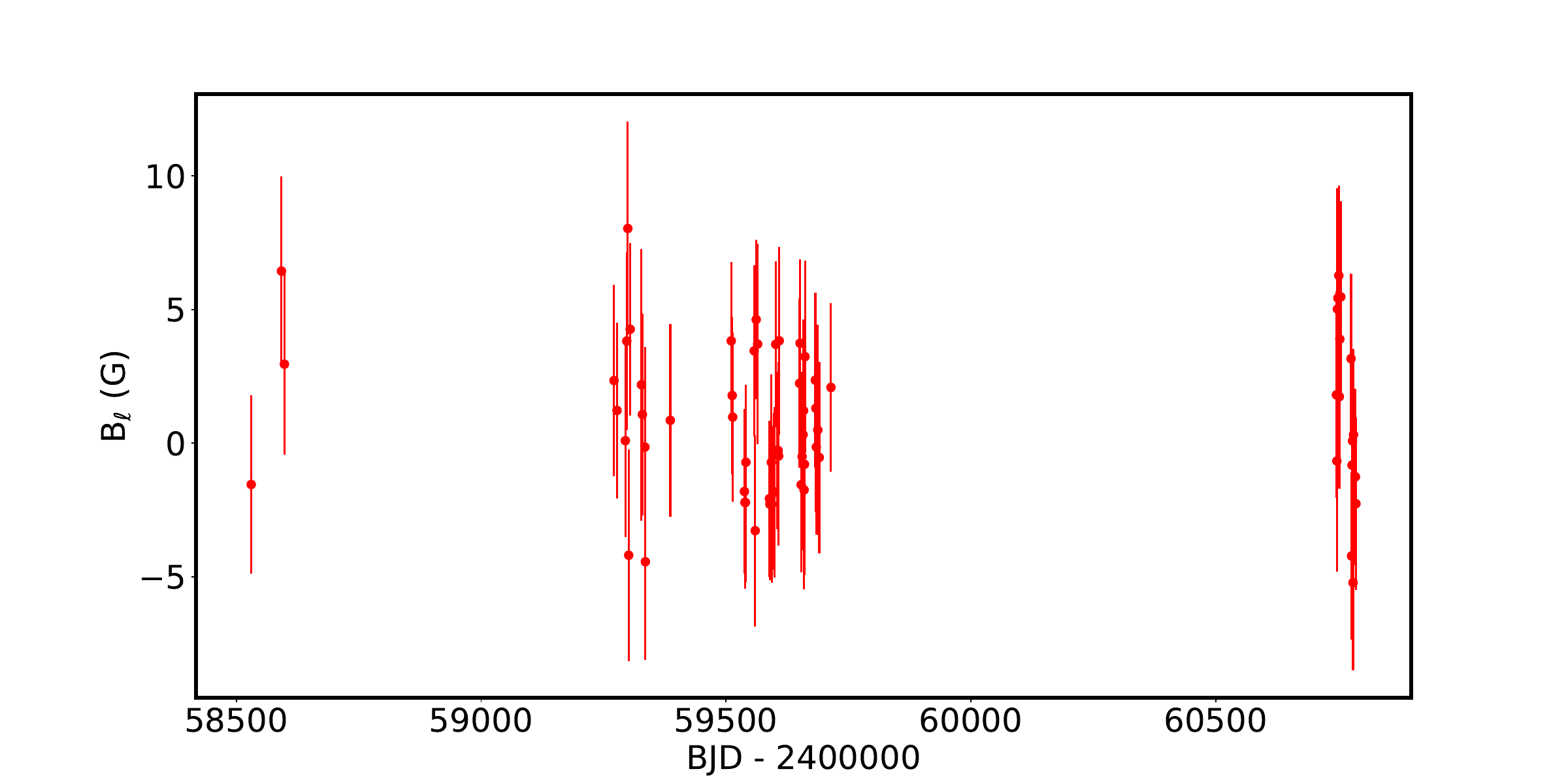}
\caption{Investigating magnetic field signatures for 55 Cnc B. Top: intensity (Stokes $I$), null (Stokes $N$), and circularly polarized (Stokes $V$) profiles. Bottom: variations of the longitudinal magnetic field as a function of time.}\label{fig:polar55CncB}
\end{figure}

Extracted and reduced with APERO, there are 70 circularly polarized spectra of 55 Cnc B. One sequence was removed due to poor quality, which were repeated trials on the same night at epoch 2460754 when two other sequences have an average quality. As for the primary star, LSD profiles were calculated from these spectra, using a mask built from MARCS models  \citep{Gustafsson2008} of a 3500 K star (keeping atomic lines deeper than 5\% of the continuum level), with 408 lines used. 
None of the Stokes V profiles shows a detected signature with a standard deviation of the LSD profile of 4$\times$10$^{-4}$ in average.
The integrated, unsigned longitudinal magnetic field in the LSD profiles is less than 10~G at 3-$\sigma$ confidence. There is no significant periodic variability in a GLS periodogram.
Figure \ref{fig:polar55CncB} shows the averaged Stokes profiles of all available 55 Cnc B polarized spectra (consistent with a longitudinal field of $0.6 \pm 0.3$~G).
In conclusion, 55 Cnc B has a low, undetected large-scale magnetic field, in agreement with its low small-scale field, and its topology cannot be characterized from this set of polarized spectra.

\section{Radial velocity analysis}

\subsection{RV data of 55 Cnc A}
The SPIRou RV time series spans 920 days and contains 141 data points. 
Note that the BERV excursion of $\pm$29.6 km/s for this star favors a good telluric correction performance \citep{ouldelhkim2026}.
Also note that the relative motion due to the binary is neglected in this analysis, due to the very large separation (orbital period of about 30000 years and semi-amplitude of about 200 m/s).

The mean SNR of used RV sequences is 510 per pixel in the $H$ band and the used exposure time is 4$\times$61.29s per polarimetric sequence. Table \ref{table:spiroudata} summarizes the average SPIRou RV data parameters.
The mean error of the SPIRou RV measurements of 55 Cnc A is 0.92~m/s. The $K$ band contributes the less to RV precision, as shown in Table \ref{table:spiroudata}. 

We then collected literature values. We used data listed in \citet{bourrier2018} obtained with various optical spectrographs: Hamilton (290), Tull (141), Keck/HIRES (24+678), Lick APF (318). We did not use public data from HRS, SOPHIE, HARPS, and HARPS-North  which were scarce compared to other data sets (mostly acquired for transit observations of planet e). We discarded 19 data points with errors greater than 5\,m/s and then binned the data into 2-h bins (sequences for transits or other phases). The total number of visits included in the analysis is 956.

\subsection{RV data of 55 Cnc B}
The SPIRou data collection spans 2250 days and contains 69 measurements, mostly obtained after February 2021. Exposure times of 4$\times$(150 to 400)s per polarimetric sequence were used, depending on the conditions. The mean SNR per pixel in the $H$ band is 180 for the sequences (Table \ref{table:spiroudata}). 
Wapiti corrections were applied to the time series and we found three principal components that need to be corrected for.

Archival RV data exist on this star: 31 measurements from Keck/HIRES (1999-2012) and 11 from CAHA/CARMENES (2016-2020),  with RMS and mean photon noise of 13.62 and 5.3 m/s for HIRES, and, respectively, 11.23 and 1.06 m/s for CARMENES \citep{butler2017, ribas2023}. Part of the SPIRou data covers the epoch of CARMENES observations. 

The relative motion of the low-mass component is expected with a semi-amplitude of 800 m/s over 30,000 years, which is negligible over the time span of the available measurements.

\subsection{Planets around 55 Cnc A}
\subsubsection{Global analysis}
The dispersion is about 50~m/s due to known exoplanet signals of large amplitude.
The RV time series of 55 Cnc A is modulated by signals of both very short and very long periods, which prevents us from using one instrument only. However, SPIRou data alone with a time span of 920 days is sufficient to independently detect planets e-b-c-f between 0.73 and 260~d periods, with a residual dispersion of 2.6~m/s or 1.7~m/s when an additional linear trend is fit to account for the second largest signal at 5000~d period.

The first analysis is performed using the tools in DACE \citep{delisle2018}, where we can easily compare various models. The first model consists of using the planet parameters in \citet{bourrier2018}, to test how this model performs with 7 more years of RV data. The RMS of the residuals is 6.0 m/s. When the magnetic cycle, fitted as a Keplerian of 3822~d period, is added, the RMS decreases to 4.5 m/s. The RMS improves to 4.0 m/s when a linear trend is added. 
This preliminary analysis shows that the literature model still stands very strongly with an additional 7 years of data collection, that a new long-term trend appears, and that the slope in the new SPIRou data is critical to constrain the longest period signals. 

The complete analysis was then performed with \radvel\ version 1.4.11 \citep{fulton2018} using the priors of \citet{bourrier2018}. Many models were compared: with and without long-term trends, and varying the priors and number of fitted signals. As in \citet{bourrier2018}, the magnetic cycle was included in the modelling as an additional Keplerian. The best model was chosen over a dozen other experiments, including 5 to 7 Keplerian models, linear and quadratic trends, and periodic or quasi-periodic modelling of the stellar cycle. 

Concerning the inclusion of stellar activity in the modelling, we face two main issues. The first is that the main activity indicator, the index derived from the Ca H and K lines, is only available in the past optical data and not in conjunction with the more recent SPIRou data. This lack of continuity requires us to extrapolate from the past cycles, as shown in Figure \ref{fig:cycle_55cncA}, either with a sinusoid or with a Gaussian process, which leads to a very imprecise result. The second difficulty is the proximity between the (quasi) period of the cycle and the period of one of the Keplerian signals, which forces us to restrain the priors. It also results in ill-constrained models at this point. Thus, it is a limitation of the present analysis, as long-term entangled signals require longer, homogeneous, combined time series. The favored model consists of 6 Keplerians (5 for the known planets, 1 for the possible magnetic cycle) as well as a linear trend.

Each experiment carried out with \radvel\ also includes a Monte Carlo Markov Chain (MCMC) statistical analysis and a Bayesian model comparison in which the significance of all signals is evaluated. The final statistical parameters of the best model are the following: MCMC chains contained 86 walkers, $7\times10^6$ steps, and $1.6\times10^6$ burning steps. The final likelihood value is $\ln{\mathcal{L}}$ = -2771. The removal of one of the 6 modelled signals leads to an increase greater than 200 in $\ln{\mathcal{L}}$. The best model is shown in Figure \ref{fig:allsignalsA} and the best fit model parameters are listed in Table \ref{tab:planetsA}.

\subsubsection{Inner system}
\begin{figure}[ht!]
    \centering
\includegraphics[width=\hsize]{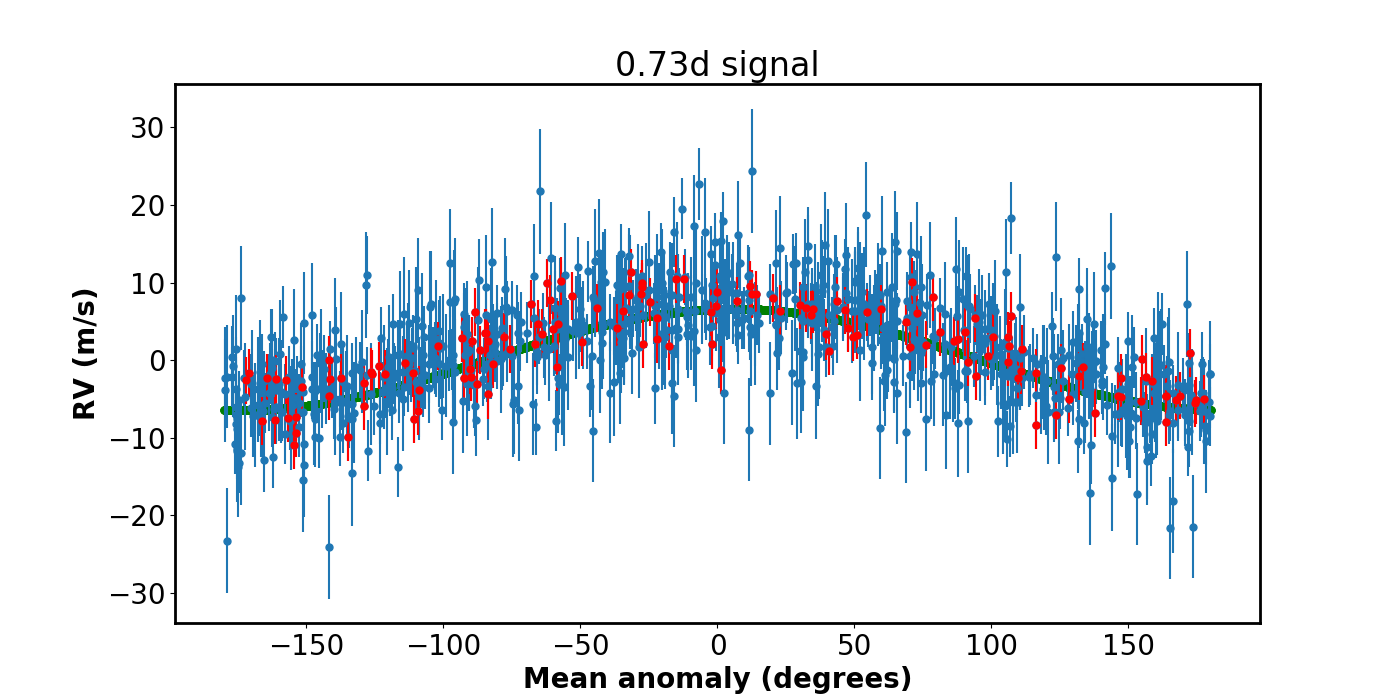}
\includegraphics[width=\hsize]{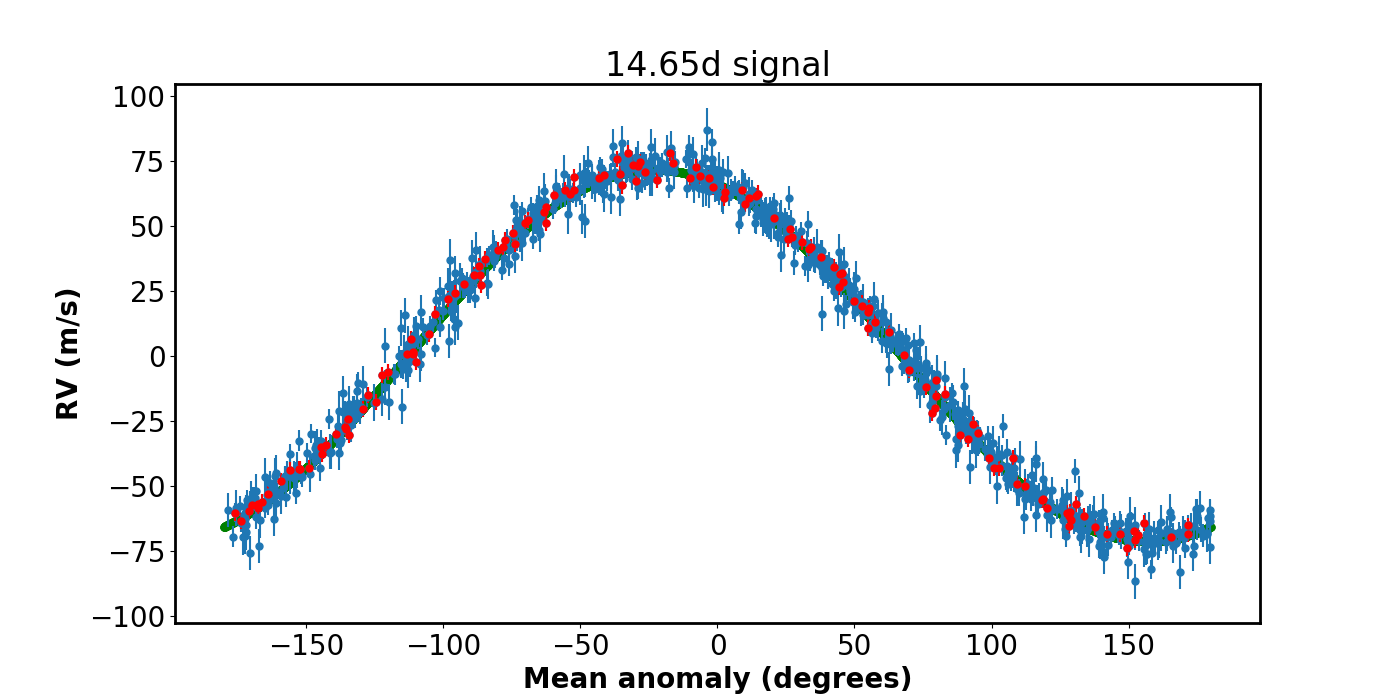}
\includegraphics[width=\hsize]{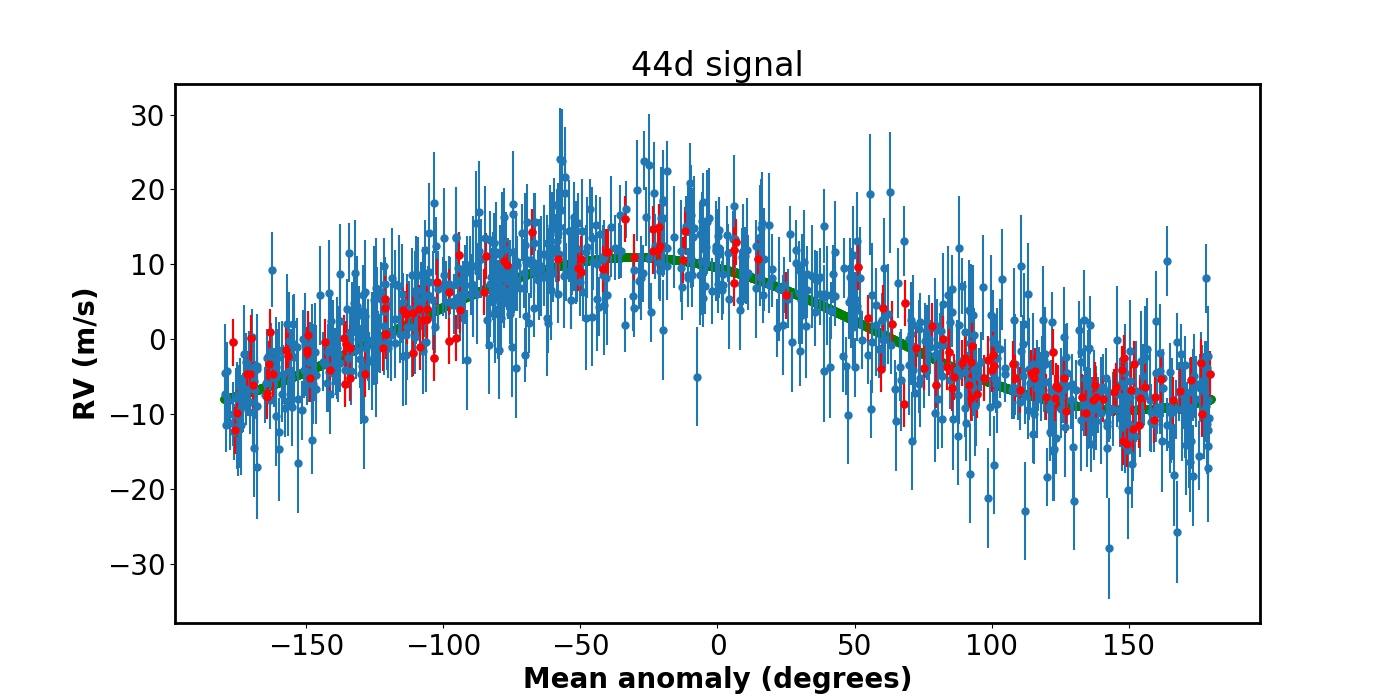}
\includegraphics[width=\hsize]{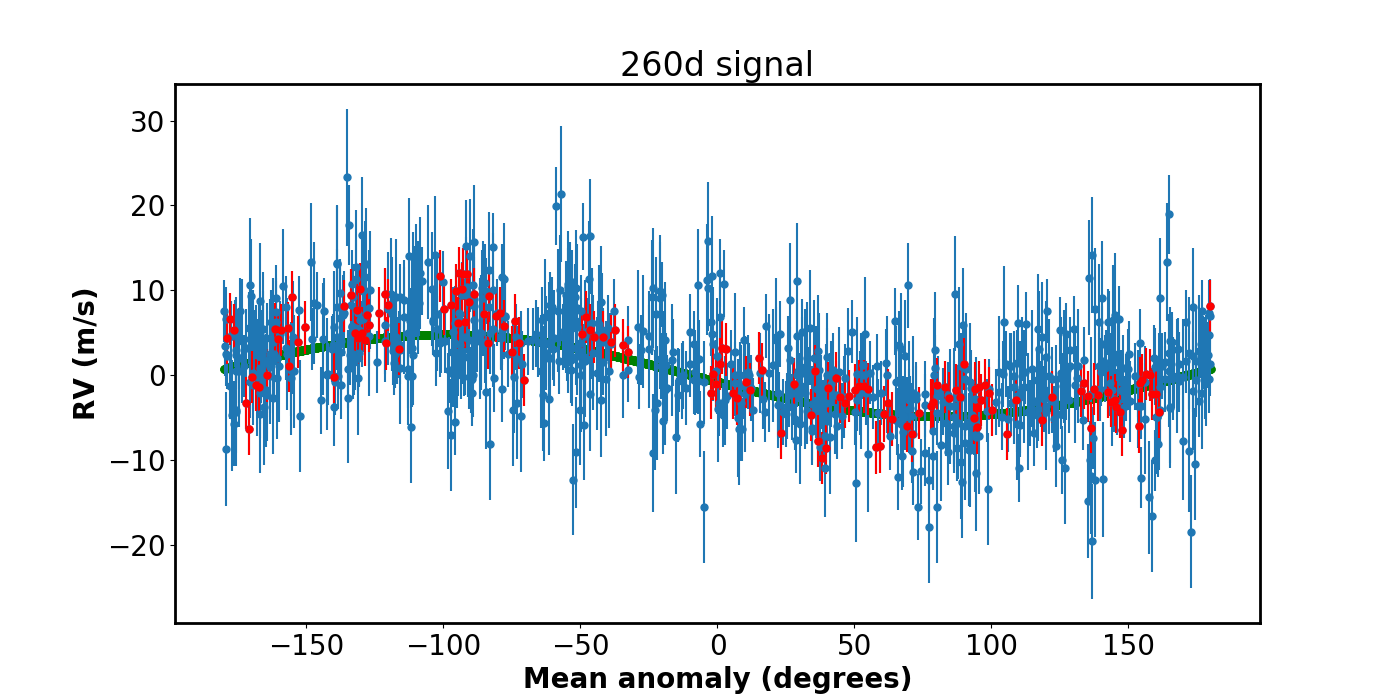}
\includegraphics[width=\hsize]{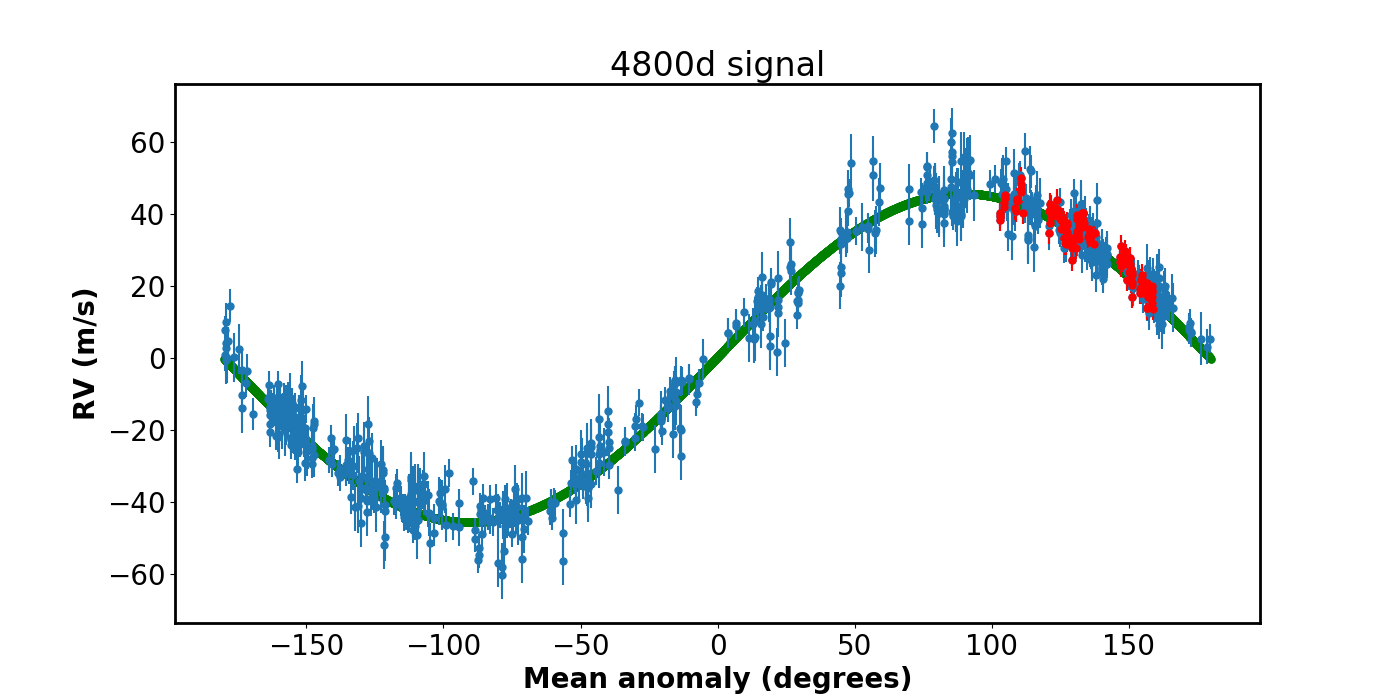}
\caption{\radvel\ analysis of 55 Cnc A: individual signals, including the 5 planets.}\label{fig:indplanetsA}
\end{figure}

The parameters of the four inner planets are remarkably stable compared to the literature values, as expected by the fact that the literature data are heavily used in this analysis and that short-period signals are less affected by the diversity of used instruments. The precision of several parameters is improved with the addition of SPIRou data, with errors on the periods typically twice that of the SPIRou data. Figure \ref{fig:indplanetsA} shows the individual RV orbits of the planets with the optical (blue symbols) and nIR/SPIRou (red symbols) RV data, illustrating the lack of chromaticity and similarity. The absence of notable surface RV jitter of this time series allows such a comparison without resorting to Gaussian processes. 

We conducted planet injection and recovery tests for a potential planet between planet f and planet d in the $O-C$ residuals. Using a period of 400 days, a planet of 8 \ME\ or more would have been detected.

\subsubsection{Outer system}
\begin{figure}
    \centering
\includegraphics[width=\hsize]{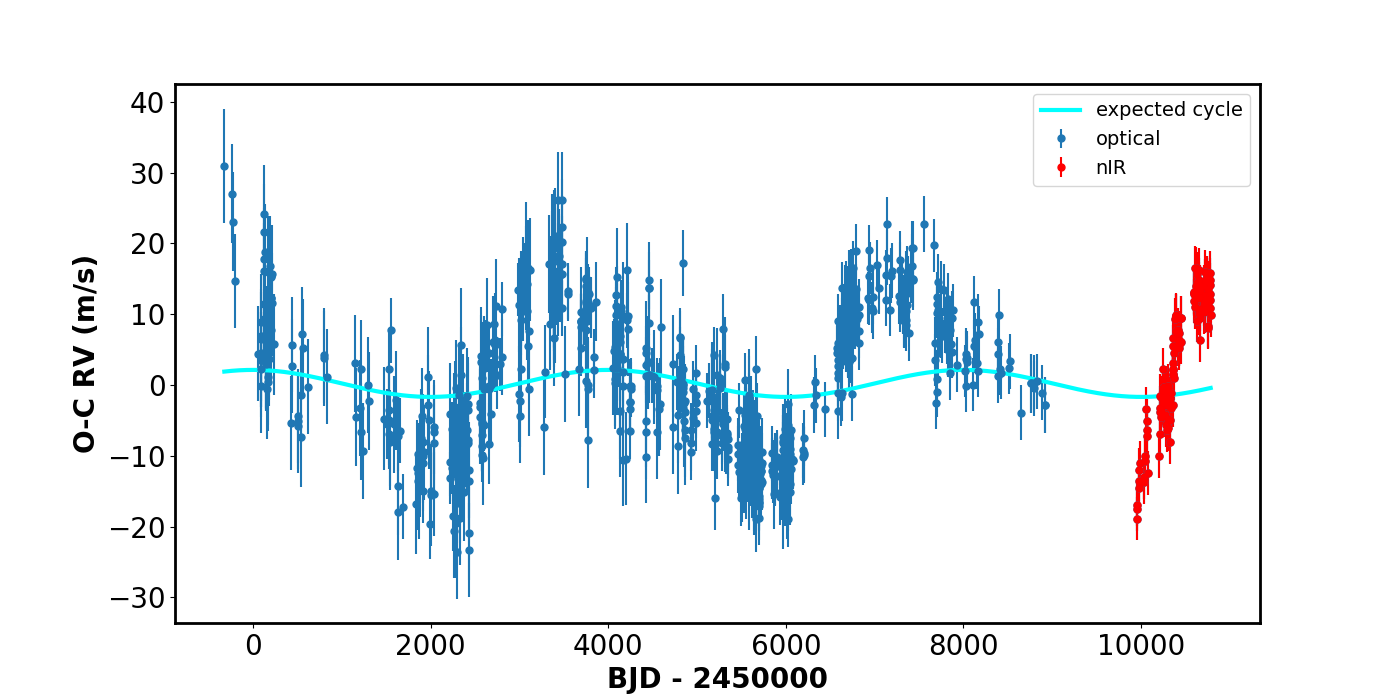}
\includegraphics[width=\hsize]{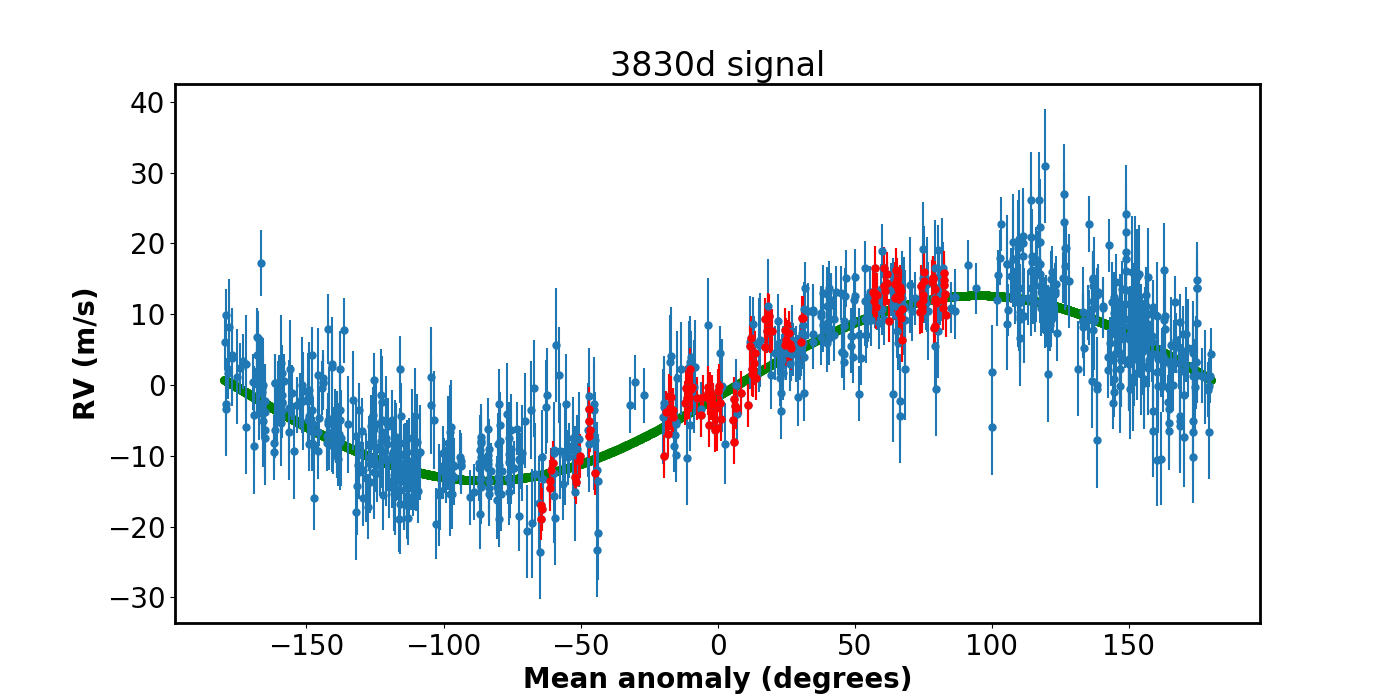}
\caption{{\it Top}: the residual RV signal of the 6th Keplerian as a function of time, when all other signals are removed. The expected model for the magnetic cycle is superimposed in cyan (see text). {\it Bottom}: The 6th Keplerian as a function of the mean anomaly with the model in green. In both figures, the blue data correspond to all optical RV measurements and the red data points to the SPIRou nIR RV measurements.   }\label{fig:planet6chrom}
\end{figure}

In the following, we revisit the assumption of \citet{bourrier2018} that one of the long-term signals is due to the magnetic cycle of the star. In this previous work, the semi-amplitude of the cycle (fitted as a Keplerian with a period of 3822$\pm77$ days) is 15.2$\pm$1.7\,m/s. Here, we found that the best-fit global model favors a $3827.1^{+9.3}_{-6.4}$ day period signal with a semi-amplitude of $13.06\pm0.48$ m/s, in agreement with the previous analysis and stable over the 28.7 years of data.
However, as was shown in the previous section, 55 Cnc A is a very quiet star: 1) the rotational modulation is barely seen, or not seen, in sensitive activity proxies despite very high SNR values (dTemp, B$_l$), 2) the large-scale magnetic field strength is 0.3 G during the cycle minimum and up to 3 G during the maximum (possibly hampered by cancellation between north and south hemispheres), and 3) the mean log $R'_{\rm HK}$ over 21 years is -5.01 (from data in \citet{isaacson2024}). This level of activity is less than the average Sun's magnetic field which has a 7 to 15 G range and a log $R'_{\rm HK}$ varying from -5.05 to -4.85 due to the 11-yr cycle, from minimum to maximum, resulting in a 2 to 6 m/s scatter \citep{Haywood2022, meunier2010,meunier2024}, mostly due to line shape changes rather than RV position variability \citep{klein2024}. In addition, \citet{meunier2010} mentions an amplitude of 8 m/s and \citet{meunier2019} predicts a 1-3 m/s scatter during the stellar cycle of a solar-type star. Other examples in which a magnetic cycle has been identified for stars similar to \hbox{55 Cnc A} (e.g., the analogs HD 1461, HD 40307, HD 204313, and HD 137388) show a cycle-induced RV full amplitude of 0.5 to 10 m/s at maximum \citep[e.g.,][]{dumusque2011,lovis2011,Diaz2016}.
Finally, we compared optical and nIR RV data over the $\sim$4000-d variation cycle of 55 Cnc A RVs. They are shown in Figure \ref{fig:planet6chrom}. The signal appears to be achromatic with in particular a steep RV increase during the SPIRou observations of about 30 m/s that matches the slope observed in the previous years with optical spectrographs. With SPIRou RVs measured in a much redder spectral range and during the $S$ index magnetic minimum, such a consistent variation and large amplitude are more consistent with a signal of planetary nature than from a magnetic cycle. The cycle, however, exists and is probably responsible for part of the RV variation at this 4000-d timescale. If ignored, the derived minimum mass might be slightly overestimated.  In Figure \ref{fig:planet6chrom}, the expected cycle signal obtained from the $S$-index sinusoidal variation is overplotted in light blue. The factor used here between log $R'_{\rm HK}$ and RV is 25 m/s/dex \citep{meunier2019}, whereas it should rather be 120 m/s/dex to match the observed amplitude at this period. In addition, the period is quite different and both cycles appear to diverge in recent years. At this point, the absence of chromaticity, the unmatching periods and the larger-than-expected amplitude question the interpretation of the 3830-d signal as being due to the magnetic cycle only, rather than mostly due to another outer planet in the system. Were the whole signal amplitude at this period due to a planet, this would correspond to a minimum mass of 0.9\,\MJ. It must be noted that other families of solutions rather include a 1900-d signal and a quadratic trend in addition to the $\sim$4000-d signal. More data will be needed to confirm the best solution. 

The outer planet d is found to have a significantly shorter orbital period of $4799.3^{+4.9}_{-12.0}$ days compared to $5574^{+94}_{-87}$ days \citep{bourrier2018} which is 9$\sigma$ longer. In past studies, the period of this planet has been found to be 4905$\pm$30 day \citep{endl2012}, 4825$\pm$39 d \citep{baluev2015}, 5285$\pm$5 d \citep{fischer2018}, and 5218 days \citep{rosenthal2021}. This apparent instability is obviously not independent from the other long-term signal - be it due to the magnetic cycle or/and a planet - in addition to the fact that the longer signals are naturally more sensitive to the use of multiple instruments than short-period ones. Finally, the best model includes a long-term trend, with a slope of 0.84$\pm$0.06 m/s/yr, possibly indicative of another outer body in the system with a period greater than 9000d (24 years). It is foreseen that more recent optical RV and $S$-index data will be published in the future, since this system attracts a lot of attention, and combined nIR-optical modelling of outer planets and a low-RV-amplitude magnetic cycle will be made possible with the aid of a continuous activity indicator.

\subsection{Planets around 55 Cnc B}
\begin{figure}[ht!]
    \centering
\includegraphics[width=0.9\hsize]{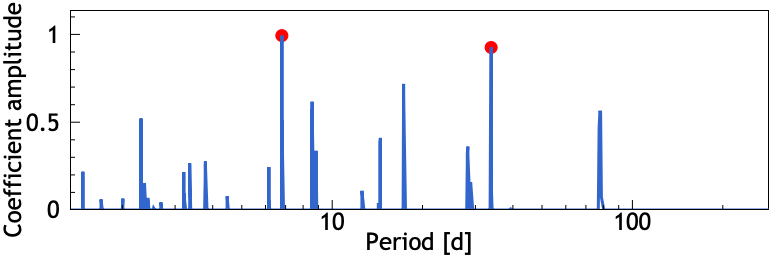}
\caption{The $\ell_1$ periodogram of 55 Cnc B SPIRou RV data, showing the prominent peaks at 6.8 and 33.7~d.
}\label{fig:rv_55cncb}
\end{figure}

\begin{figure}[ht!]
    \centering
\includegraphics[width=\hsize]{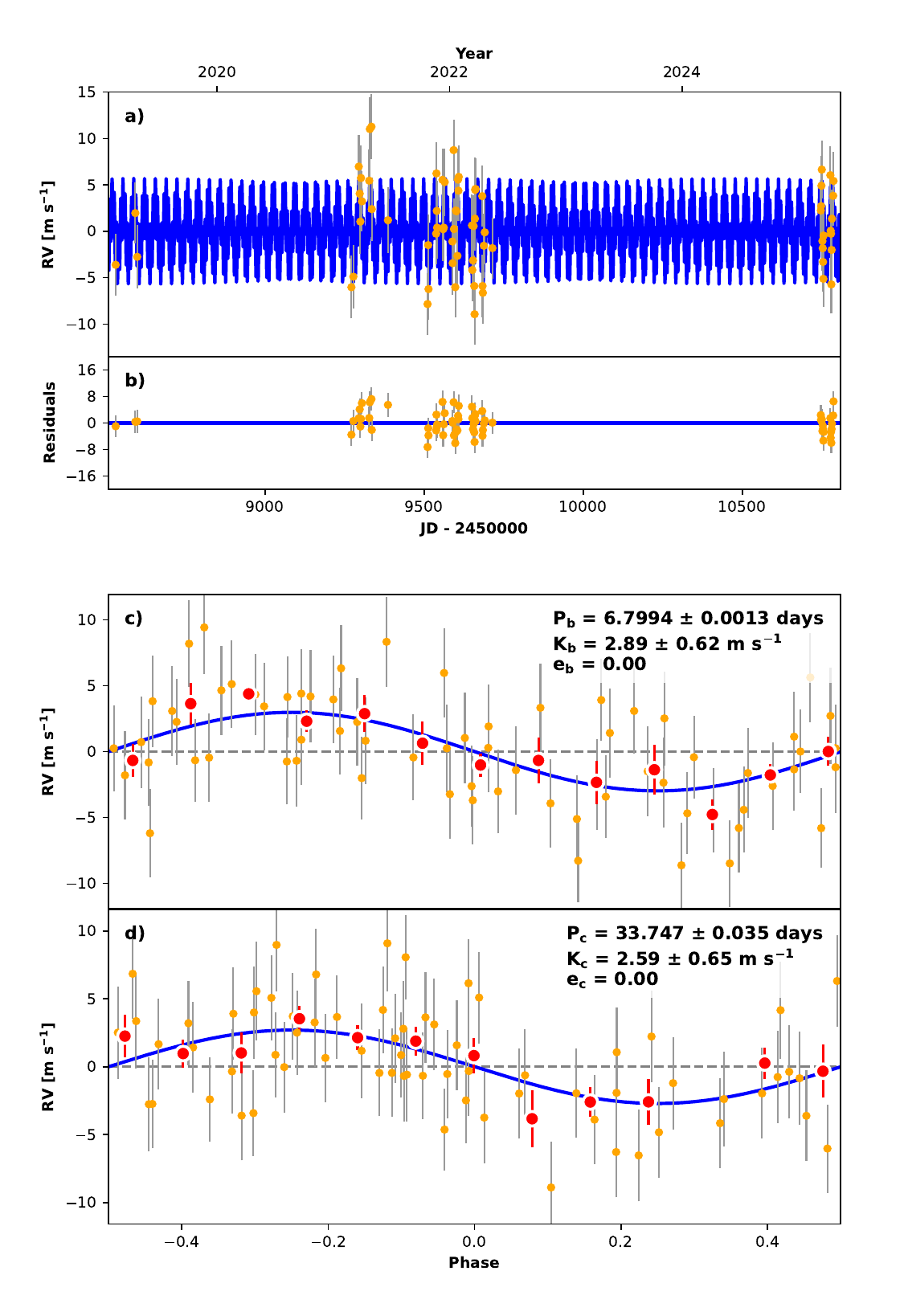}
\caption{The orbital solution found from SPIRou ("s") RV data of 55 Cnc B as a function of time (top) and phase (middle and bottom for planets b and c, respectively). The residuals are shown in panel b. Red points are mean measurements per regular 0.1 phase bins as calculated with \radvel\ when enough data is available in the bin.}
\label{fig:55cncb_rv_spirou}
\end{figure}

We have analyzed the 69 RV measurements obtained with SPIRou on 55 Cnc B.
SPIRou RV measurements of 55 Cnc B have a mean error of 1.8~m/s and a dispersion of 5.9~m/s. The $\ell_1$ periodogram \citep{hara2017} shows a prominent peak at 6.8d with a $logBF$\footnote{$BF$ stands for the Bayes Factor, defined as the differential Bayes Information Criterion as in \citet{ouldelhkim2025})} of 10.3 and a $logFAP$ \footnote{FAP stands for the False Alarm Probability, which encodes the probability that a given peak in the periodogram be due to noise rather than to a periodic signal} of -0.0015 (Figure \ref{fig:rv_55cncb}). Another peak is also present at 33.7\,d, with a lower probability.
Both periods are different from the rotation period of 92$\pm$5\,d (Table \ref{tab:GPR_dt_55CNCAB}) or first harmonics of it. The 6.8d and potentially 33.7d signals could thus be due to planetary companions. We then used \radvel\ to derive the orbital parameters and run a MCMC analysis with one-planet and two-planet models.
The best-fit model is shown in Figure \ref{fig:55cncb_rv_spirou} together with the residuals. The signal is best modelled with the two-planet model in circular orbits with periods of $6.799\pm 0.002$ days and $33.747\pm0.035$\,d and semi-amplitudes of $2.89\pm0.63$ m/s and $2.59\pm0.66$\,m/s, respectively. \radvel\ finds a $\Delta$AICc\footnote{$\Delta$AICc is the relative Akaike Information Criterion
as defined in https://radvel.readthedocs.io/} improvement of 604 with the two-planet model with respect to none, and a preference of the two-planet model over either one-planet model. The planets causing these RV modulations have a minimum mass of 3.5$\pm$0.8 and \hbox{5.3$\pm$1.4\,\ME} and a semi-major axis of 0.045$\pm$0.001 and 0.130$\pm$0.003\,au, respectively. When the two-planet model is removed, we find no other detected signal in the residuals, including at the rotation period, and the RMS of the residuals is 3.35\,m/s. Interestingly, the period ratio is close to 5, which explains the beating pattern in Figure \ref{fig:55cncb_rv_spirou}. However, this is too great a ratio for an active resonance. All parameters for the 55 Cnc B planets are summarized in Table \ref{tab:planetB}.

The available archival data are either too noisy or too scarce to detect the signal seen in SPIRou data. For example, when injecting the 6.8\,d signal 1000 times into the precise but scarce CARMENES data, it was retrieved only 21 times with a FAP smaller than 0.1\% (0 times with the HIRES data, and 145 times with both). The complete model is shown in Figure \ref{fig:allsignalsB}.

\begin{table}
\caption{Fitted parameters for planets b and c of 55 Cnc B  \label{tab:planetB}}
\begin{tabular}{lcc}
\hline\hline
Parameter & Prior & Posterior  \\\hline
P$_b$ (d) & $\mathcal{N}$(6.8,0.5) & $6.7994^{+0.0013}_{-0.0014}$   \\
Tc$_b$ & $\mathcal{N}$(2455495.6,1)&  $2455495.7^{+0.84}_{-0.85}$   \\
K$_b$ (m/s) &$\mathcal{J}$(1,10)& $2.89^{+0.61}_{-0.63}$  \\
e$_b$ & fixed & 0 \\
P$_c$ (d) & $\mathcal{N}$(33.7,0.5)& $33.747\pm 0.035$  \\
Tc$_c$ & $\mathcal{N}$(2455492,5)& $2455491.8^{+4.1}_{-4.2}$   \\
K$_c$ (m/s) &$\mathcal{J}$(1,10)& $2.59^{+0.65}_{-0.66}$  \\
e$_c$ & fixed& 0  \\
$\sigma_{\rm SPIRou}$ (m/s) &$\mathcal{N}$(3,3) & $2.91^{+0.34}_{-0.31}$  \\\hline
m$_b$sin$i$ (\ME) & -& 3.5$\pm$ 0.8   \\
a$_b$ (au) & -& 0.044$\pm$0.0012  \\
m$_c$sin$i$ (\ME) & -& 5.3$\pm$1.4   \\
a$_c$ (au) & -& 0.13$\pm$0.003\\\hline
\end{tabular}
\tablefoot{$\mathcal{J}$ ($\mathcal{N}$) stands for a Jeffreys (resp., normal) prior}
\end{table}

\begin{figure}
    \centering
\includegraphics[width=\hsize]{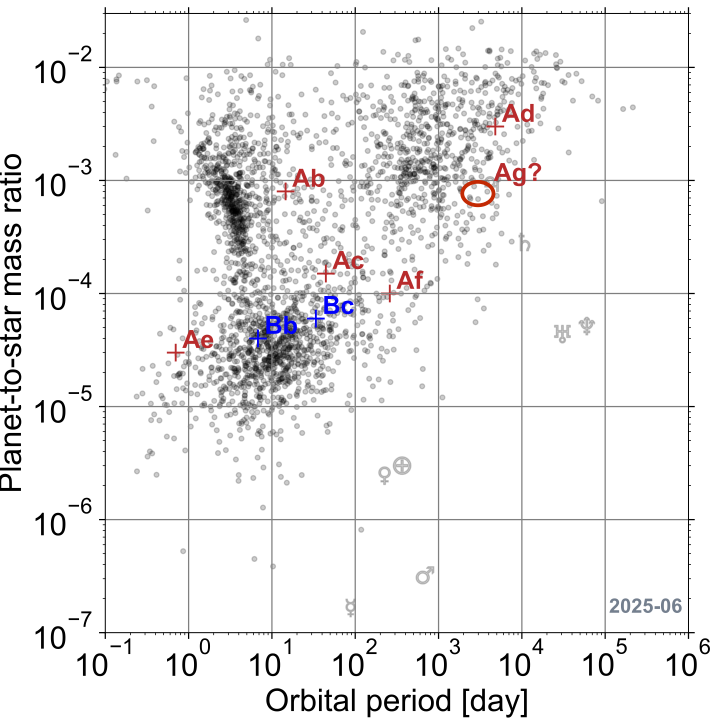}
\caption{Planet-to-star mass ratio as a function of the orbital period for all known exoplanets. The planets orbiting both components of 55 Cnc are highlighted. Symbol "Ag?" corresponds to the maximum mass of a planet at 3830-d period, if it gets confirmed in the future.}
\label{fig:ccl}
\end{figure}

\section{Discussion}
\subsection{Habitable zones}
From this new RV analysis, no new RV signal is found in between the inner set of four planets and the outer planet(s) of 55 Cnc A with a 10 year period or more. The outer part of the habitable zone is still unpopulated at this time. A planet with more than $\sim$ 8 \ME\ in a circular 400-d period orbit (or semi-major axis of about 1 au) would have been detected in our data set, although there will always be suspicion of a tentative detection near 1-yr orbital periods. The eccentricity of planet f being 0.063$\pm$0.048 in our analysis, slightly smaller than found in \citet{bourrier2018}, the dynamical stability of a potential temperate Earth-like planet between 1 and 2 au in the system is confirmed and reinforced \citep{satyal2019}. Although challenging, the detection of a 1-3 \ME\ exoplanet at 1-1.5 au in this complex data set may be envisioned because of the quiet surface of the star, but will require some more years of regular observing, both in optical and nIR. Note that planet f is also located in the habitable zone of its star in between the runaway and maximum greenhouse limits \citep{Hill2022}, with a minimum mass of about 48.5\,\ME.

Concerning the two planets in the 55 Cnc B system, we find that the first planet has an insolation of 2.6 S$_\oplus$ and an equilibrium temperature of 330-370~K, and the second one has an insolation of 0.3 S$_\oplus$ and an equilibrium temperature of \hbox{190-210~K}, assuming an albedo in the range of 0.50-0.25, respectively. It is safe to say that 55 Cnc Bc lies near the outer edge of the habitable zone of its star, receiving about as much energy as Mars from the Sun. Thus, it seems that the lower mass secondary star of the 55 Cnc system also has at least a temperate planet in its habitable zone extending from approximately 0.055 to 0.14\,au. 
The 55 Cnc system may be the first binary with one planet in each habitable zone.

\subsection{Planets in binary systems}
 55 Cnc is only the sixth binary system (out of 739 planet-hosting binaries known to date) in which both stellar components are known to host a planetary system, together with WASP-94 (A and B), Kepler-132 (A and B), XO-2 (N and S), HD20782 and HD20781, and HD113131 (A and B) \footnote{Information retrieved from the "planets in binaries" database \citep{thebault2025}: https://exoplanet.eu/planets$\_$binary/}. We stress that this low number is very likely to be strongly biased by the fact that, for a large fraction of planet-hosting binaries, no dedicated investigation of planets around the companion star has been performed. However, it is worth pointing out that among these six systems, 55 Cnc stands out because of the very low mass ratio (0.27) between the two stellar components, while all other 5 systems have mass ratios close to 1. In fact, 55 Cnc B is one of the smallest known planet-hosting stars that reside in a binary. 

\begin{figure}
    \centering
\includegraphics[width=\hsize]{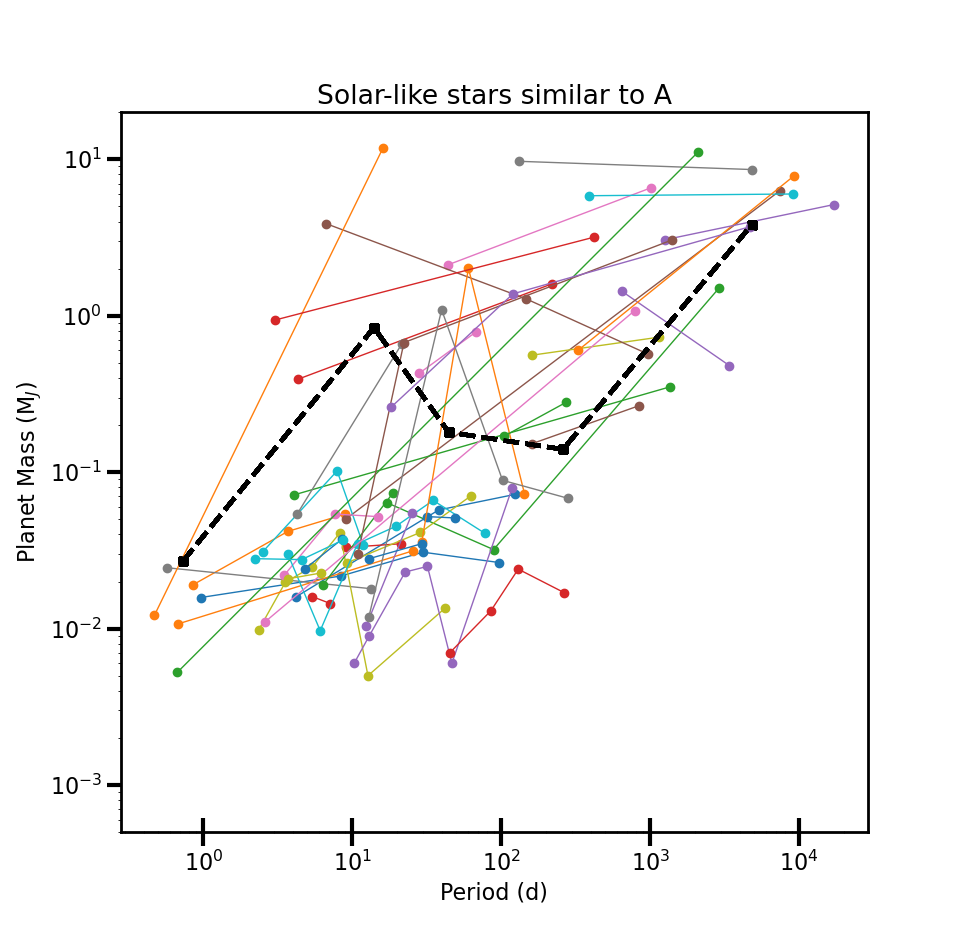}
\includegraphics[width=\hsize]{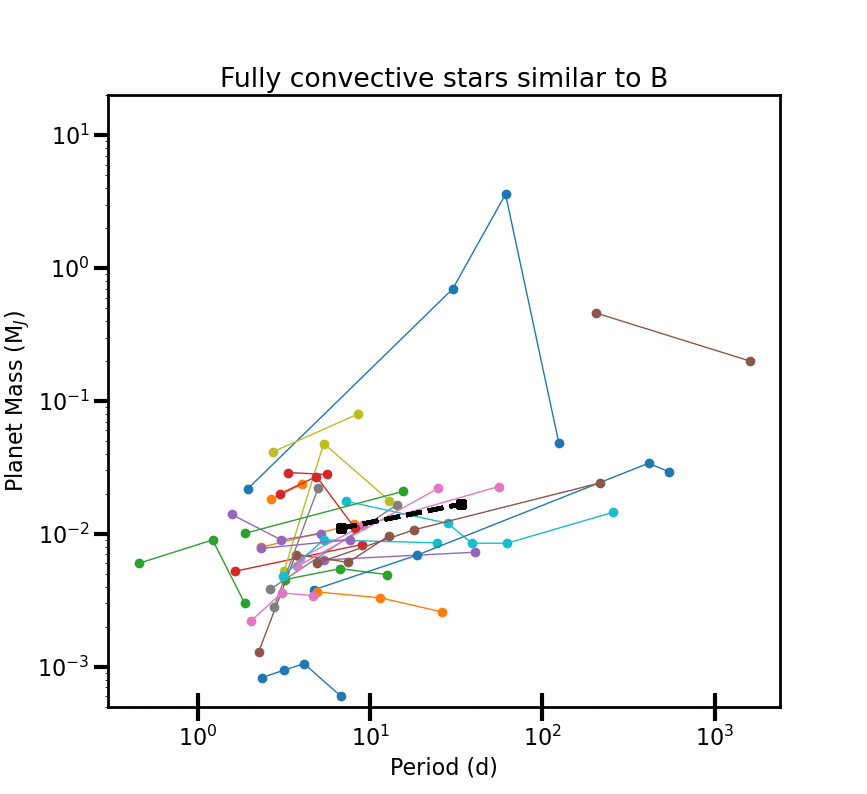}
\caption{Projected mass and orbital period of multiple systems around stars similar to 55 Cnc A (top) and 55 Cnc B (bottom). Symbols for the same system are connected by a line. The black dashed lines show the confirmed planets of 55 Cnc stars. Data are obtained from the Exoplanet Encyclopedia at https://exoplanet.eu/. }
\label{fig:mul}
\end{figure}

Figure \ref{fig:ccl} presents the planet-to-star mass ratio versus orbital period of the confirmed planets and potential candidate (marked by a "Ag?" and a large ellipse) in the 55 Cnc system compared to other systems. While the two planets of the secondary join a populated region of this diagram with numerous inner super-Earths, the planets of the primary lie in less populated locations.
Comparing the 55 Cnc multiple systems to other systems around similar stars as in Figure \ref{fig:mul} illustrates the large observed difference: systems around fully-convective stars (bottom plot, with stellar masses less than 0.35\,\MS, 27 systems) are in a vast majority composed of low-mass planets whose mass increases with orbital period, as previously noticed by \citet{weiss2018}. The system of 55 Cnc B matches this behaviour. On the other hand, systems around solar-like stars (top plot, 45 systems) have larger contrasts in the planet masses, often include an outer giant planet, and sometimes include a giant planet in between less massive planets.

\section{Conclusion}
We have conducted a comprehensive analysis of the 55 Cnc system, composed before this article of two stars and five planets. We used data acquired with CFHT/SPIRou over several years and were able to study the radial velocity and the circular polarization in stellar lines for both stars. 

\paragraph{The magnetic properties of the primary} 
The primary star is an old Sun analogue and has a rotation period of 39$\pm$1 days. The magnetic signal of 55 Cnc A has dropped in amplitude in the early 2020s due to the expected cycle minimum, compared to a previous campaign that was carried out during the cycle maximum in 2017, with about a factor ten in amplitude. We may expect the star to invert its polarity near the next maximum of the cycle, which will happen in 2029, if its behaviour is solar like \citep{charbonneau2020}. The updated cycle length is $\sim$4000 days or 10.96 years, coincidently close to the solar cycle. 
The factor of ten in the large-scale field amplitude results in a similar factor in the stellar wind intensity and mass loss. MHD simulations of space weather around 55 Cnc A were conducted by \citet{folsom2020} at cycle maximum, and their outputs can be scaled down by typically an order of magnitude at cycle minimum. Moreover, the small-scale magnetic field of the star during the SPIRou campaign, at minimum, is on average 250$\pm$20 G. 

\paragraph{The magnetic properties of the secondary} 
The rotation period of 55 Cnc B was estimated to be 92$\pm$5\,d from modulations of the differential temperature in the SPIRou data.
Observed for the first time in high-resolution spectropolarimetry, the magnetic field of the secondary 55 Cnc B appears to be weak compared to other slowly-rotating low-mass stars \citep[e.g.,][]{lehmann2024}. It is characterized by a small-scale unsigned amplitude less than 60~G and a large-scale amplitude less than 10~G (3$\sigma$ upper limit). With no detection in the nIR, it was not possible to find any modulation or derive its most probable topology. These upper limits are supportive of a quiet stellar surface, favourable to exoplanet searches with the radial velocity method.

\paragraph{The exoplanet system around 55 Cnc A} 
The four inner planets (periods from 0.73 to 260 days) are confirmed, and their parameters are reassessed by this analysis. The errors on their orbital parameters are improved with the addition of new data. In particular, the eccentricity of the inner planet b is found to be smaller than previously \citep{nelson2014,bourrier2018} and is compatible with 0 in the 1.1$\sigma$ range. The new value adds evidence to the tidally damped dynamics of the orbit, characterized by rapid circularization and the absence of a 3:2 spin-orbit resonance as discussed in \citet{ferrazmello2025}. \\
There is more ambiguity about the outer system of this star. The period of the giant planet is now best modelled with an orbital period of 4800 days and a mass of 3.8 \MJ. This orbit is only slightly eccentric or compatible with the circular one (eccentricity of 0.09$\pm$0.07). Another signal at 3827$\pm$9 days is also seen with a RV semi-amplitude of 13.1$\pm$0.5 m/s. This signal is in phase with the magnetic cycle as observed in the $S$ index, at least during the first recorded cycle between 1994 and 2017. There are indications that the RV signal at this period is less in phase with the $S$ index over the last decade, and also that the amplitude of this signal may be too large to be attributed to the magnetic activity. In addition, the amplitude is not chromatic and the star is extremely quiet. The possibility thus exists that part of the RV signal at 4000-d period (typically a few m/s) is due to the cycle and the other part to an outer planet at a similar period. In the solar system, the Jupiter orbit (4330 days) and the solar cycle (4015 days) are also coincidently very close. Only more data, both optical and nIR, and spectropolarimetry, will allow disentangling RV magnetic cycle signatures and potential planetary signal in the long term. The potential confirmation of a second outer planet will require in-depth dynamical stability analyses to be carried over, as strong interactions with the massive outer planet are expected. Moreover, there is still the possibility that the best-fit orbital periods of these outer planets evolve with additional data, as seen previously for planet d.

\paragraph{The first planets around 55 Cnc B}
The presence of exoplanets around the secondary component was expected for several reasons: 1) common formation conditions with the primary, 2) the system is metal rich (although B is less metal-rich than A), 3) low-mass stars in general are formed with several low-mass planets \citep[e.g.,][]{dressing2015,burn2021}. However, because of its fainter luminosity, finding these planets with the RV method requires long exposures in the optical or the use of nIR spectrographs. SPIRou is well adapted to such stars and was able to find two first signals that we interpret as being caused by exoplanets, with 69 epochs of RV measurements. The most prominent short-period planet in the system has a 6.8\,d period and a minimum mass of \hbox{3.5\,\ME}. It is located at a semi-major axis of 0.044$\pm$0.001\,au and is best modelled with a circular orbit at the moment. The second planet is characterized by a larger minimum mass (5.3 \ME) and an orbital period about 5 times longer. This pair of planets has a 5:1 period ratio, leaving room for additional low-mass planets in the system. More data would help strengthening the detection of the 33.7-d signal of 55 Cnc B, as this signal is detected at only 3.7$\sigma$ in the present data. However, the presence of giant planets in the 55 Cnc B system is excluded by the data at this stage, up to orbital periods of a couple of years. The "small sister" of \hbox{55 Cnc A} has its own planetary system, characterized by a more modest size and total mass than its primary. 

Spectropolarimetry in both optical and nIR spectral domains would be interesting to compare, as the physical processes responsible for the cycle may have different chromatic behaviour in both amplitude and scatter. This has never been studied in any star, not even in the Sun. This combined optical-nIR analysis will soon be possible with instruments like VISION at T\'elescope Bernard Lyot and Wenaokeao at CFHT which will preserve the polarimetric information while giving a simultaneous access to the relative spectrographs covering 370 to 2450 nm in one shot.

\begin{acknowledgements}
We are grateful to the referee for the careful reading and insightful suggestions that greatly improved the manuscript.
\par
This project received funding from the European Research Council under the H2020 research \& innovation program (grant 740651 NewWorlds).
\par
We acknowledge the funding from Agence Nationale pour la Recherche (ANR, project ANR-18-CE31-0019 SPlaSH) and from the Investissements d’Avenir program (ANR-15-IDEX-02), through the “Origin of Life” project of the Universit\'e Grenoble Alpes.
\par
This publication makes use of The Data \& Analysis Center for Exoplanets (DACE), which is a facility based at the University of Geneva (CH) dedicated to extrasolar planets data visualisation, exchange and analysis. DACE is a platform of the Swiss National Centre of Competence in Research (NCCR) PlanetS, federating the Swiss expertise in Exoplanet research. The DACE platform is available at https://dace.unige.ch.
\par
This research has made use of the SIMBAD database,
operated at CDS, Strasbourg, France \citep{Wenger2000}.
\par
This work has made use of the VALD database, operated at Uppsala University, the Institute of Astronomy RAS in Moscow, and the University of Vienna.
\end{acknowledgements}

\bibliographystyle{aa}
\bibliography{00_references}

\begin{appendix} 

\section{Table of data and parameters}
The full tables of SPIRou data are available in electronic format at CDS. Exerts are shown in Tables \ref{tab:dataA} and \ref{tab:dataB}. Table \ref{tab:polarbaseA} lists the optical spectropolarimetric data found in Polarbase for 55 Cnc A, also available electronically in its full length. 

Table \ref{tab:planetsA} lists the complete set of derived parameters for the planetary system of the primary star.

\begin{table}
\centering
\caption{SPIRou data of 55 Cnc A.}
\label{tab:dataA}
\begin{tabular}{lcccc}
\hline
JD-2450000& RV& $\sigma(RV)$ & B$_\ell$ & $\sigma(B_\ell)$\\\hline
 & (m/s) & (m/s) & (G) & (G) \\
9956.130208 &27945.547 &0.798 &0.6& 0.5\\
9957.136976 &27966.722 &0.912 &0.5& 0.7\\
9958.101103 &27994.811 &0.837 &-0.1&0.6\\
9978.041783 &27986.682 &0.969 &0.0&0.5\\
... & ... & ... & ... & ...\\\hline
\end{tabular}
\end{table}

\begin{table}
\centering
\caption{SPIRou data of 55 Cnc B.}
\label{tab:dataB}
\begin{tabular}{lcccc}
\hline
JD-2450000& RV& $\sigma(RV)$ & B$_\ell$ & $\sigma(B_\ell)$\\
 & (m/s) & (m/s) & (G) & (G) \\
\hline
8531.01447 & -4.196& 1.854& -1.54 & 3.33\\
8592.79288 &  1.380& 1.804 & 6.44 &3.53\\
8598.77900 & -3.337& 2.019& 2.95 &3.39\\
9271.83780 & -6.608& 1.95& 2.34& 3.58\\
... & ... & ... & ... & ...\\\hline
\end{tabular}
\end{table}

\begin{table}
\centering
\caption{Polarbase data of 55 Cnc A.}
\label{tab:polarbaseA}
\begin{tabular}{lcc}
\hline
 JD-2450000   &  B$_\ell$  &   $\sigma(B_\ell)$\\\hline
 7841.937390 &       2.13 &       0.40 \\ 
 7801.402170 &       5.84 &       0.95 \\ 
 7798.514100 &       5.96 &       1.02 \\ 
 8150.039570 &      -0.03 &       0.55 \\ 
 ... & ... & ...\\\hline
\end{tabular}
\end{table}

\section{Corner plots}
\begin{figure*}[ht!]
    \centering
\includegraphics[width=0.9\hsize]{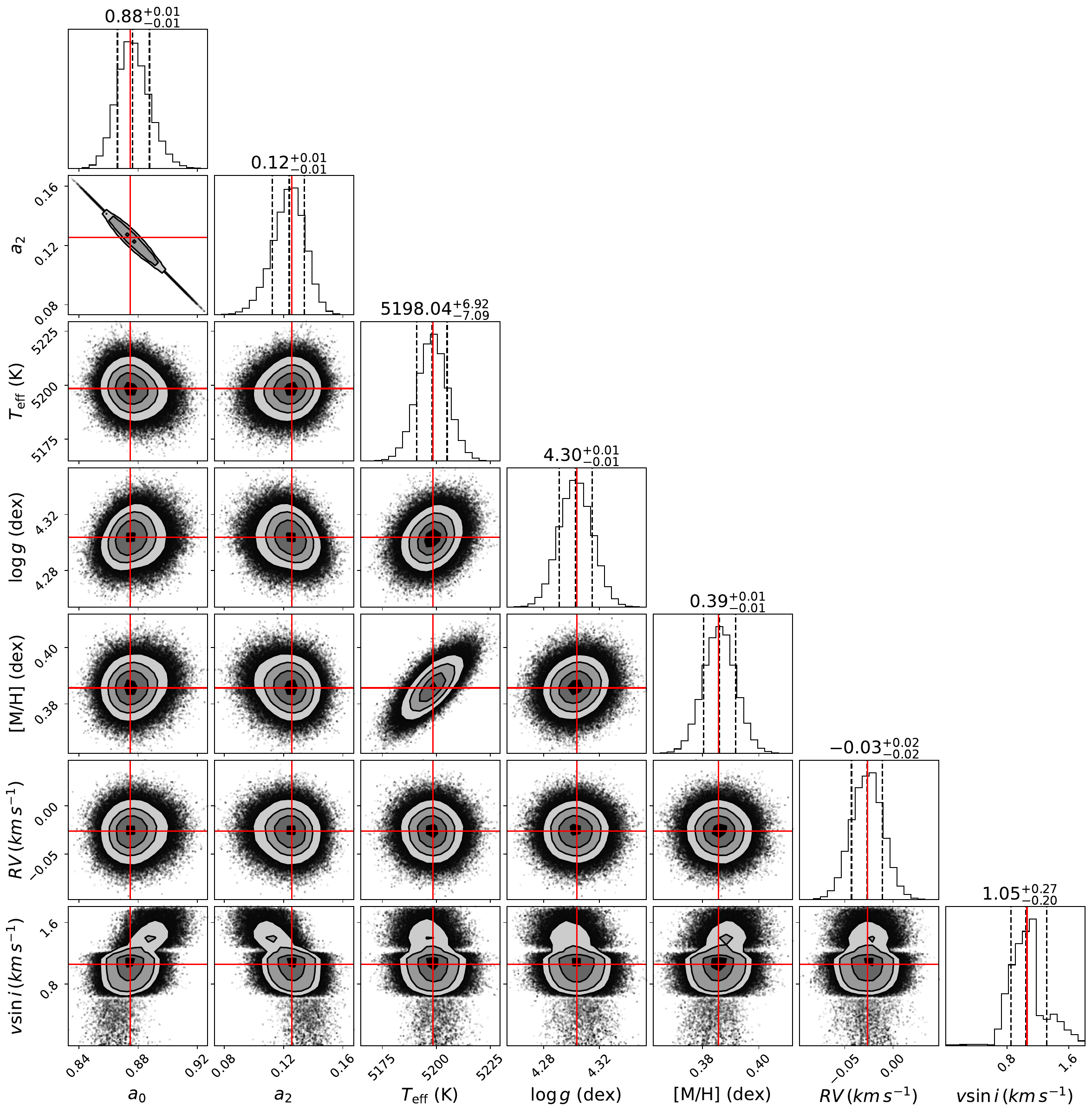}
\caption{Posterior distribution of the fundamental parameters of 55 Cnc A as derived with SPIRou: effective temperature, log$g$, [M/H], RV, vsin$i$, and the filling factors of the 0 and 2 kG surface small-scale magnetic field, a$_0$ and a$_2$. }\label{fig:corner55cncA}
\end{figure*}

\begin{figure*}[ht!]
    \centering
\includegraphics[width=\hsize]{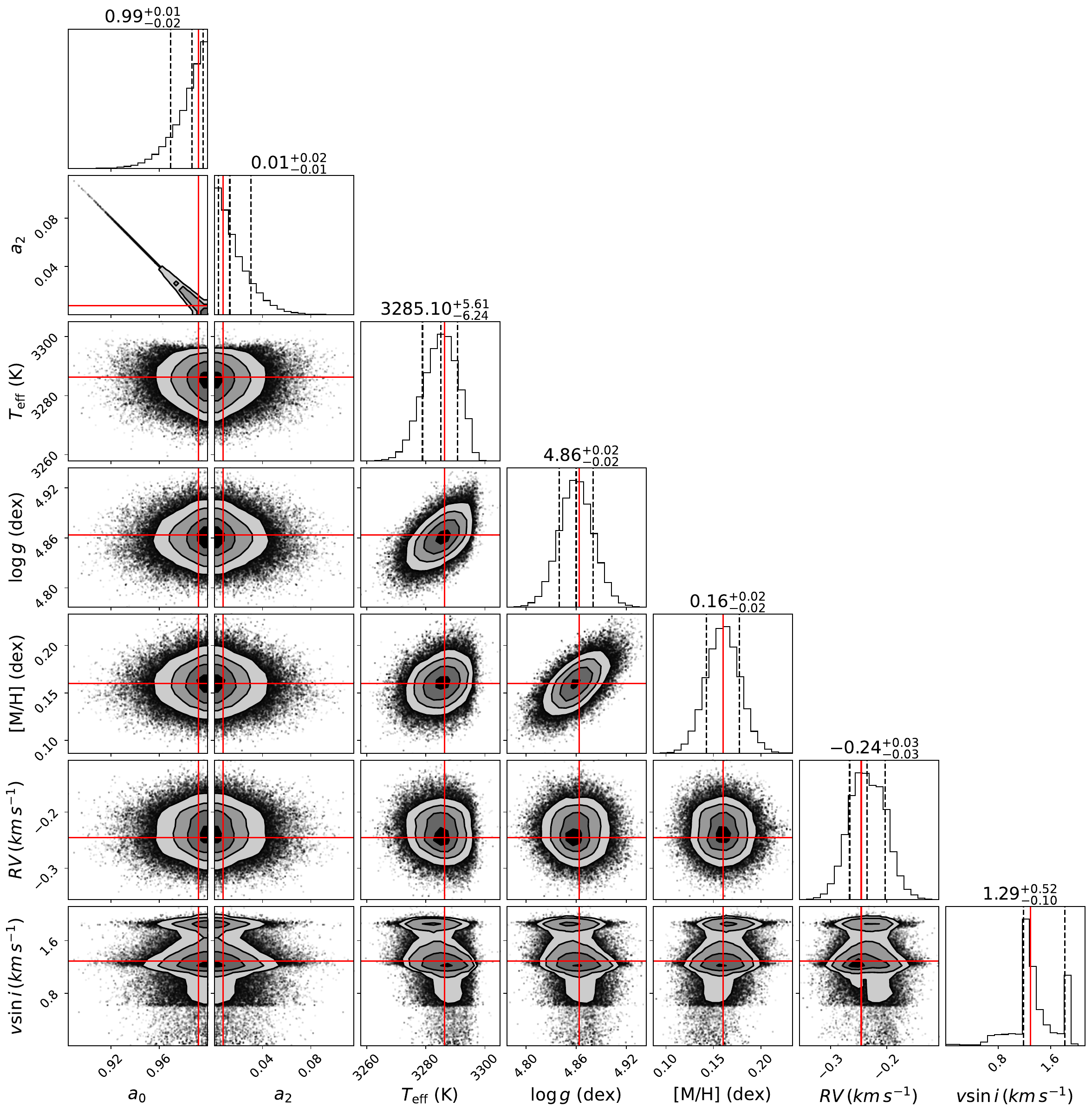}
\caption{Same than Figure \ref{fig:corner55cncA} for 55 Cnc B. }\label{fig:corner55cncB}
\end{figure*}

In this Appendix, the corner plots of the fits computed in this article are shown. Figures \ref{fig:corner55cncA} and \ref{fig:corner55cncB}  show the fundamental stellar parameters obtained from the SPIRou templates of 55 Cnc A and B, as modelled with $Zeeturbo$.

\begin{figure*}[ht!]
    \centering
\includegraphics[width=0.9\hsize]{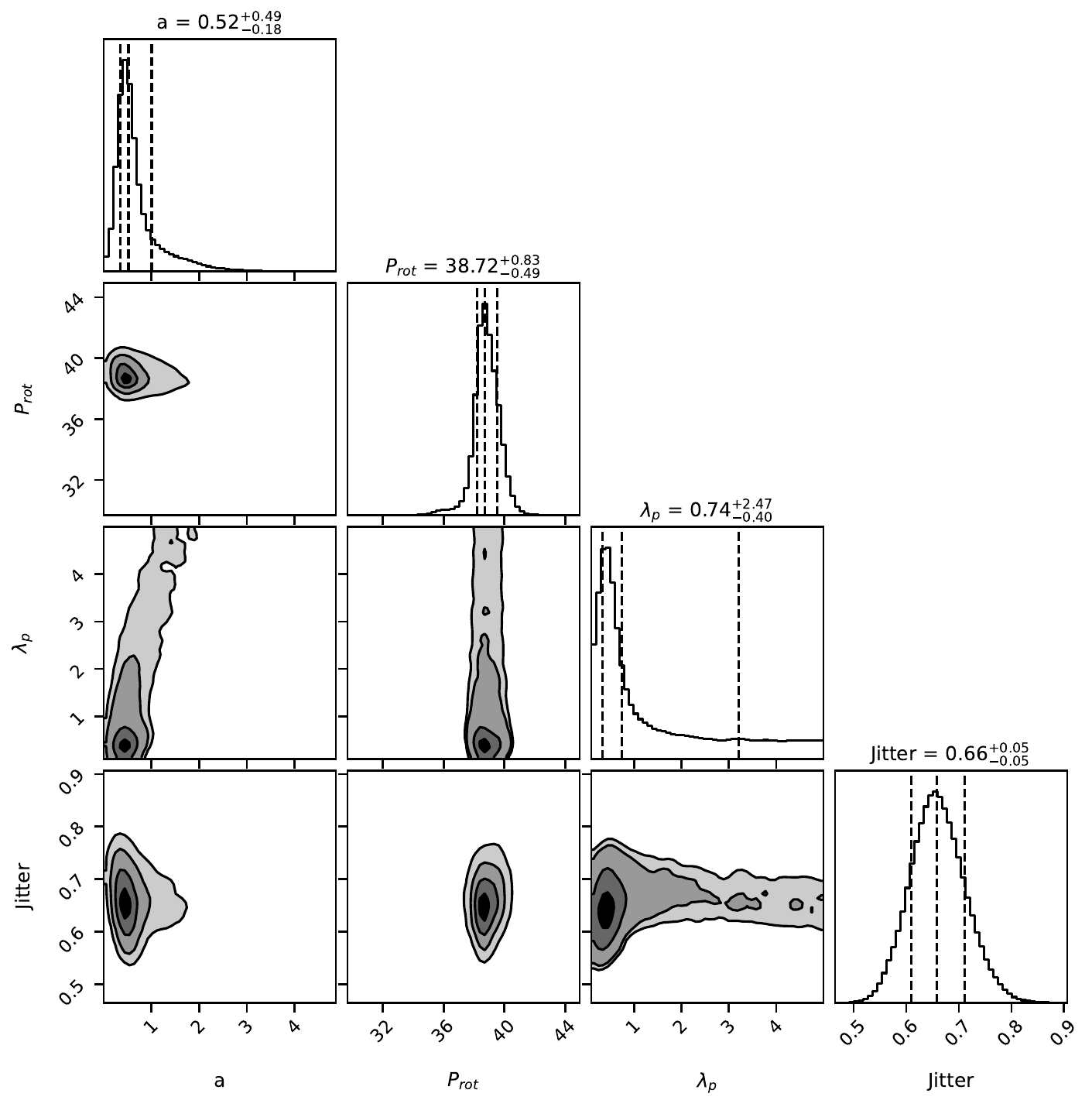}
\caption{Corner plot of the quasi-periodic Gaussian Process modelling for dTemp of 55 Cnc A: amplitude of temperature variations in K ($\eta_1$), rotation period ($\eta_3$), smoothing parameter ($\eta_4$), and jitter. }\label{fig:app1_dtA}
\end{figure*}

\begin{figure*}[ht!]
    \centering
\includegraphics[width=0.9\hsize]{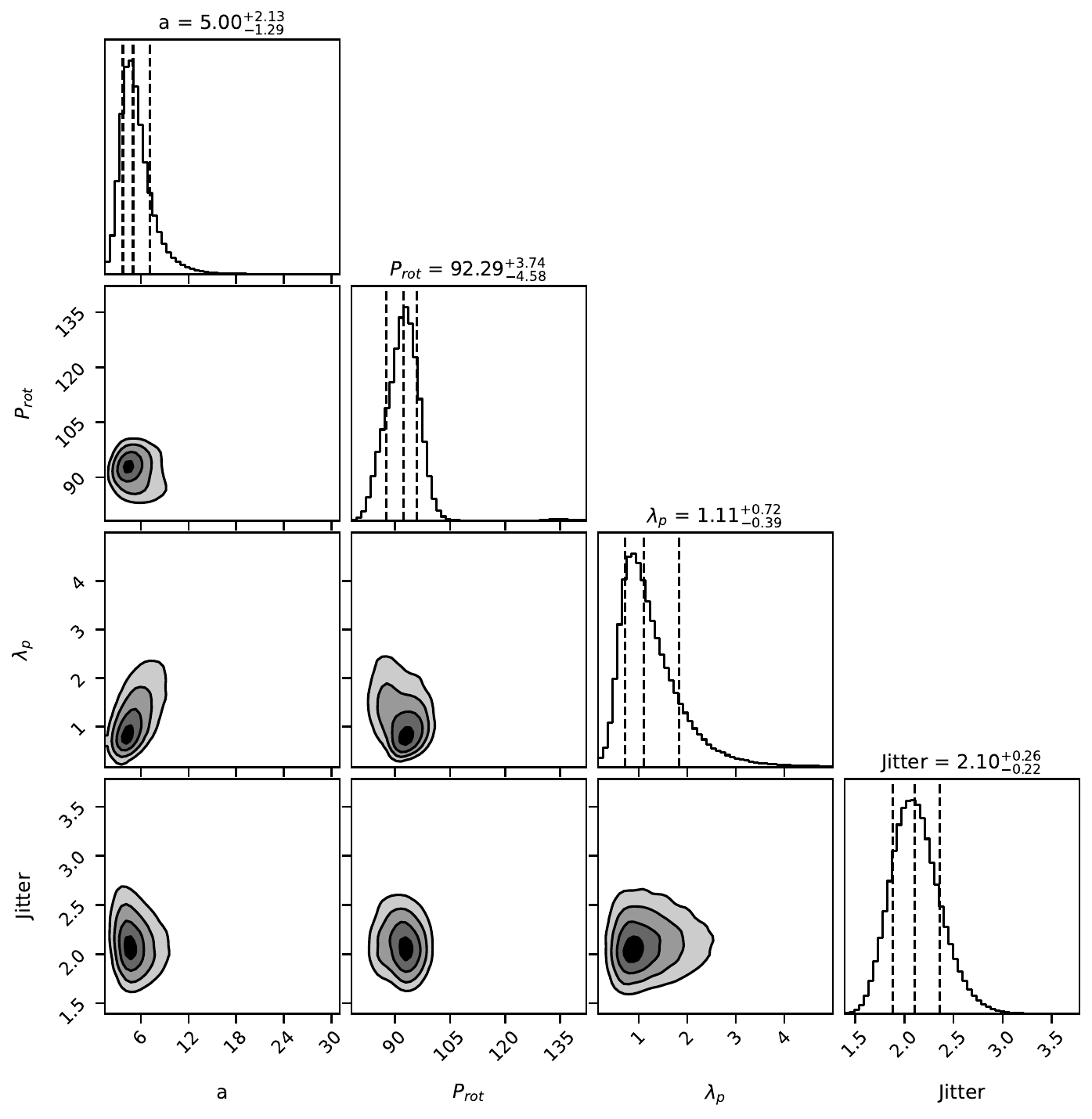}
\caption{Same than Figure \ref{fig:app1_dtA} for 55 Cnc B. }\label{fig:app1_dtB}
\end{figure*}

Figures \ref{fig:app1_dtA} and \ref{fig:app1_dtB} show the parameters of the time series of the differential effective temperature estimated in SPIRou spectra for 55 Cnc A and B, respectively.

\section{Full RV models}

\begin{figure*}
    \centering
\includegraphics[width=0.8\hsize]{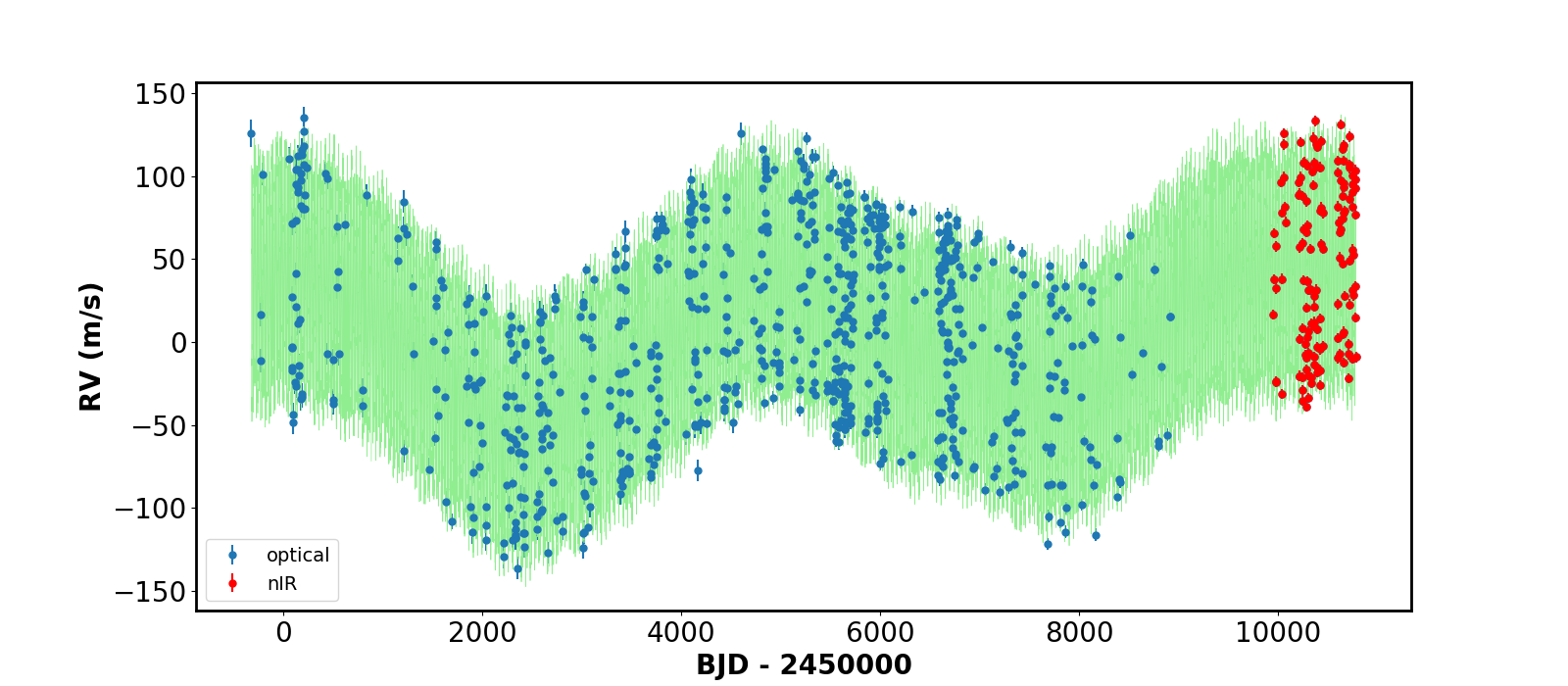}
\includegraphics[width=0.8\hsize]{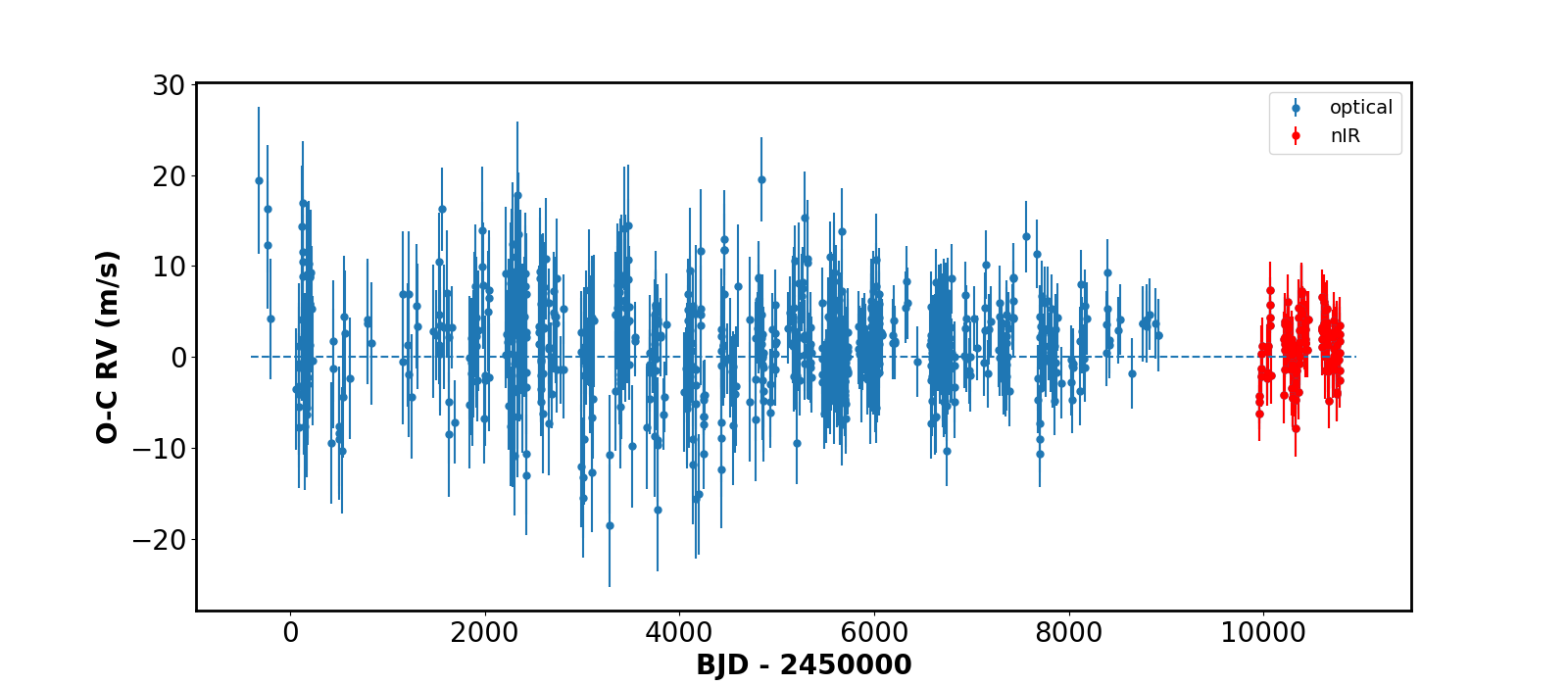}
\includegraphics[width=0.8\hsize]{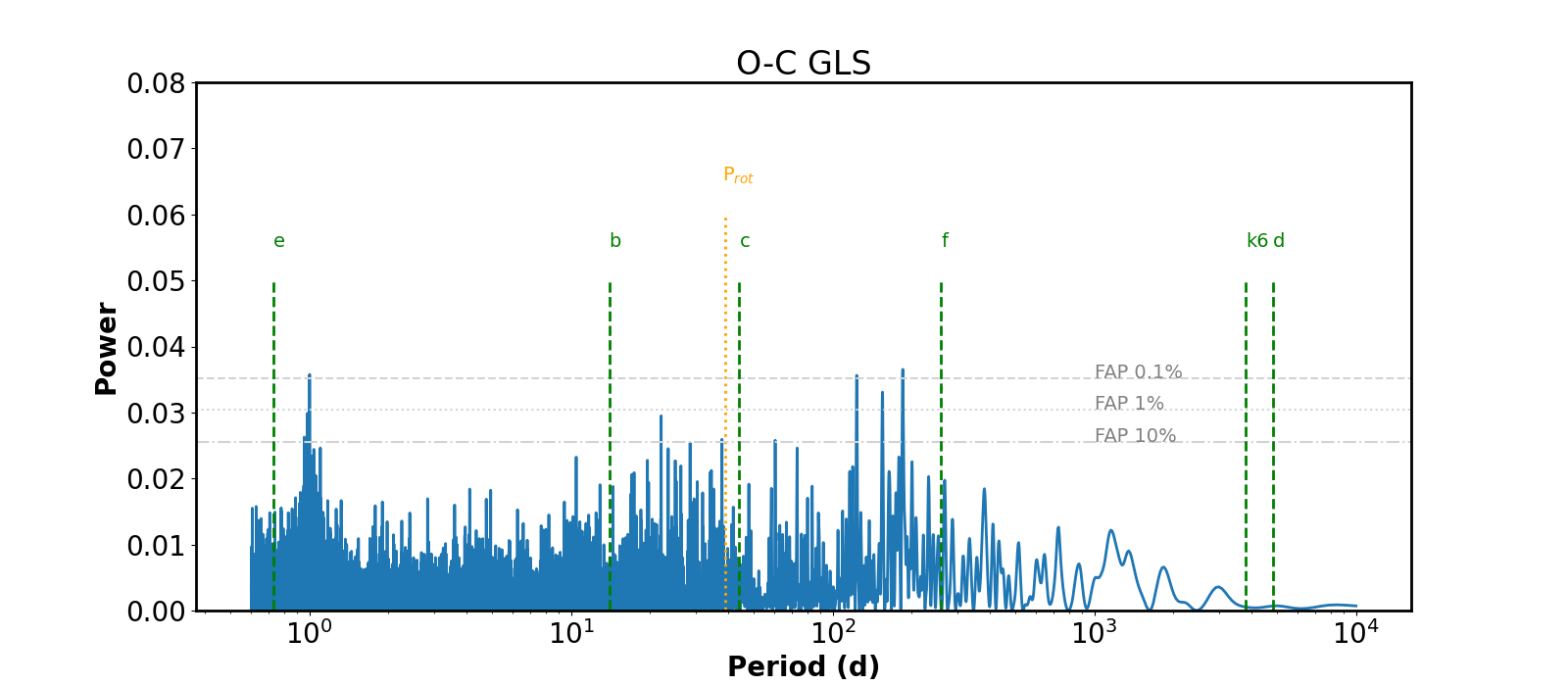}
\caption{\radvel\ analysis of 55 Cnc A: (top) all instruments (optical instruments in blue, SPIRou in red) and all signals, as a function of time, (middle) $O-C$ residuals, (bottom) GLS periodogram of residuals.}\label{fig:allsignalsA}
\end{figure*}

Figure \ref{fig:allsignalsA} shows all RV data used in the analysis of 55 Cnc A, with blue symbols for all optical data and red symbols for nIR SPIRou data. The raw RVs are shown in the top plot, the O-C residuals are shown in the middle plot after subtraction of 6 Keplerian signals, and the bottom plot shows the periodogram of the residuals with the modelled orbital periods and the rotation period.

\begin{figure*}
    \centering
\includegraphics[width=0.8\hsize]{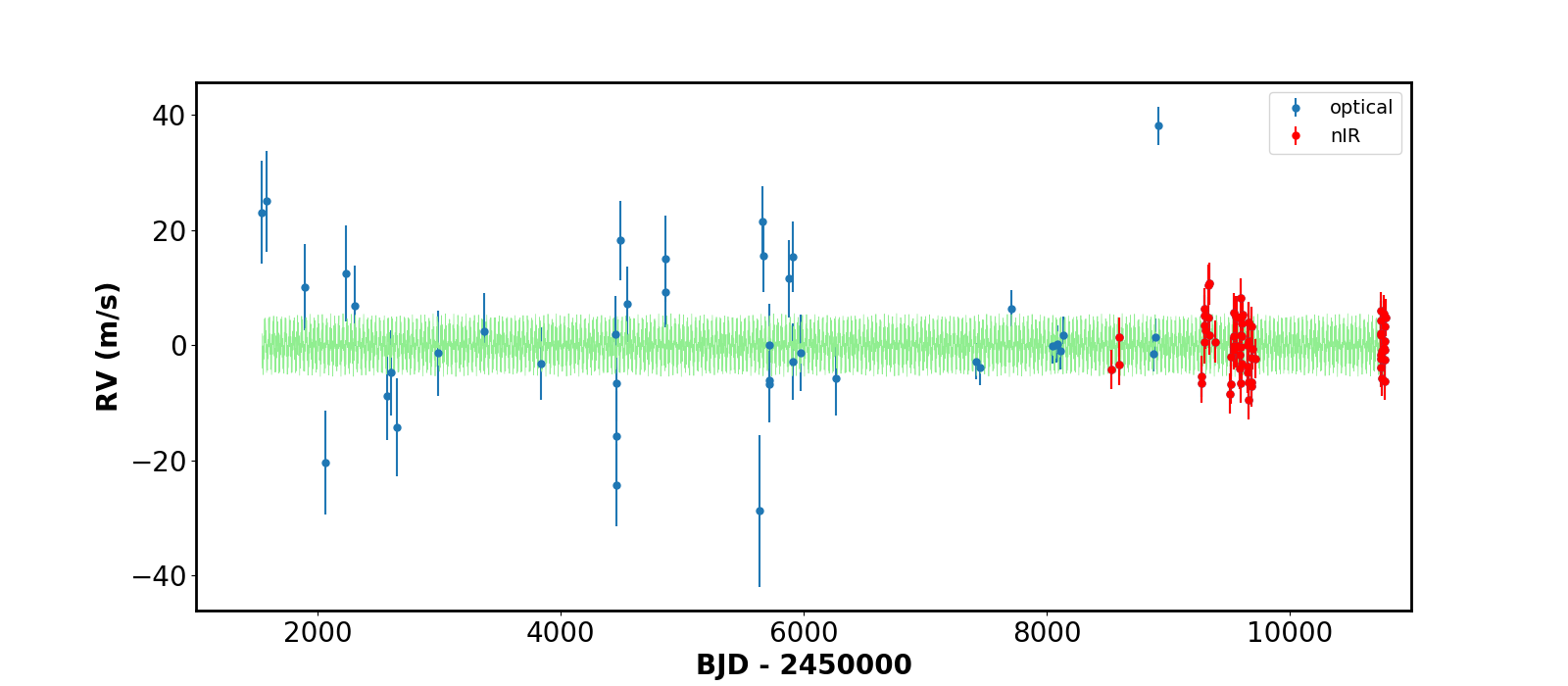}
\includegraphics[width=0.8\hsize]{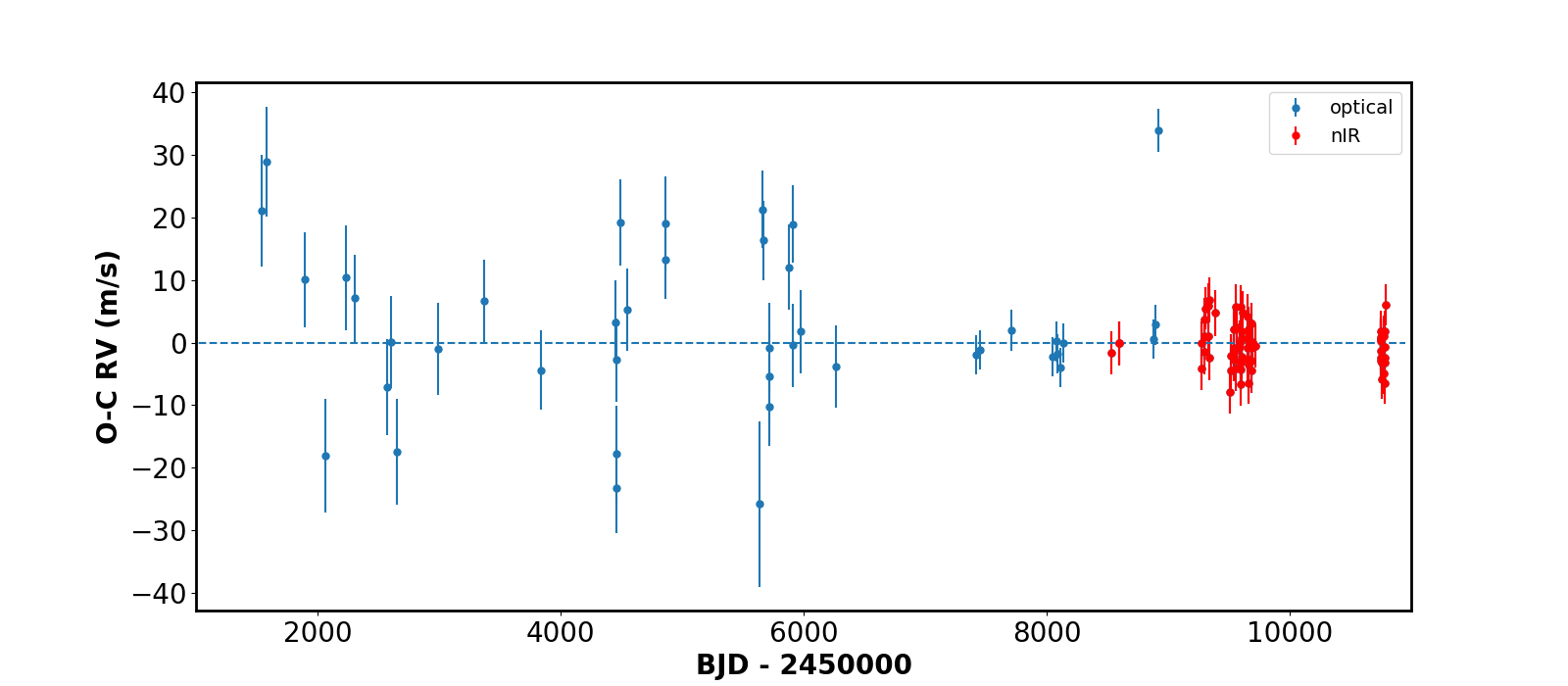}
\includegraphics[width=0.8\hsize]{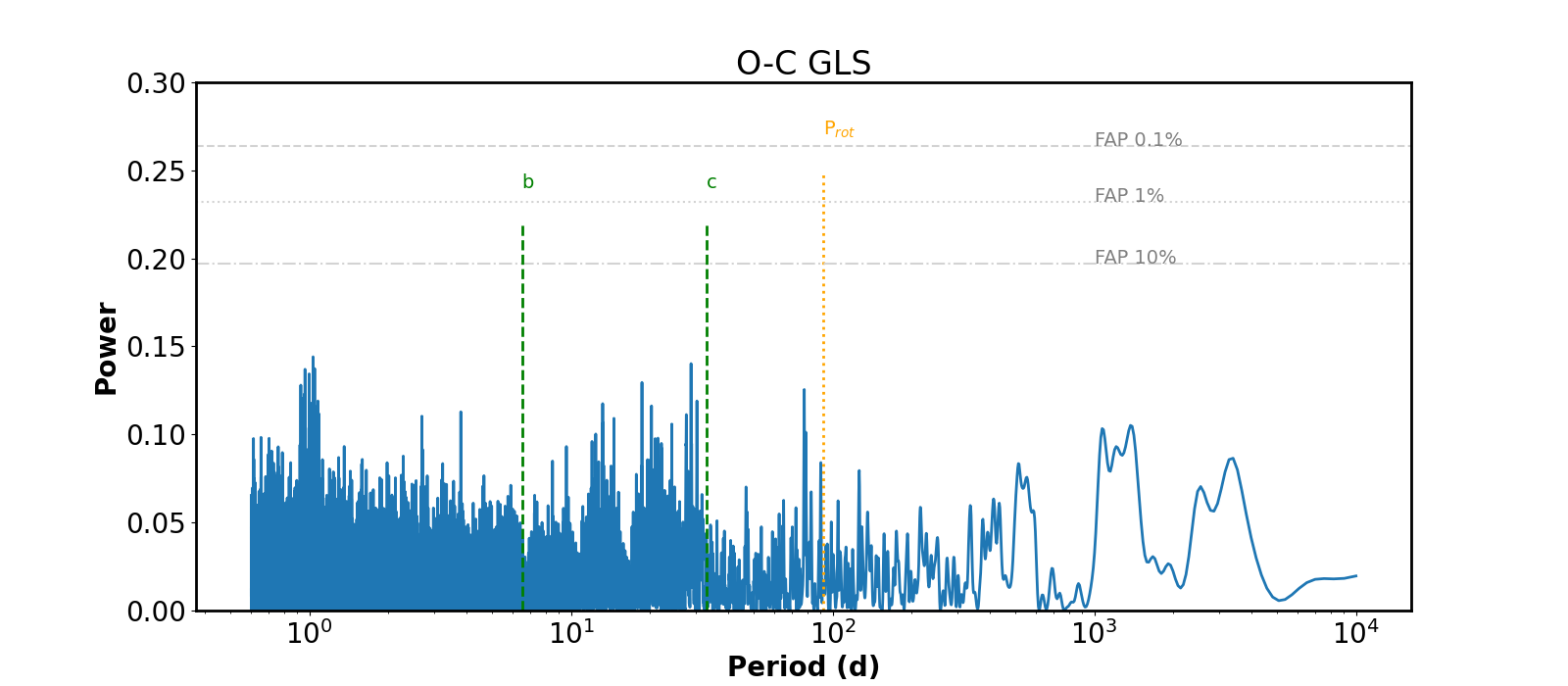}
\caption{\radvel\ analysis of 55 Cnc B: (top) all instruments (HIRES and CARMENES in blue, SPIRou in red) and all signals, as a function of time, (middle) $O-C$ residuals, (bottom) GLS periodogram of residuals. On the top panel, the model is calculated using SPIRou data alone, and overplotted over all data.}\label{fig:allsignalsB}
\end{figure*}

Figure \ref{fig:allsignalsB} shows the full model for the RV time series of 55 Cnc B, including HIRES and CARMENES data (blue symbols) and SPIRou data used for the fit (red symbols). It also shows the residuals and its periodogram.

\begin{figure*}[ht!]
    \centering
\includegraphics[width=0.9\hsize]{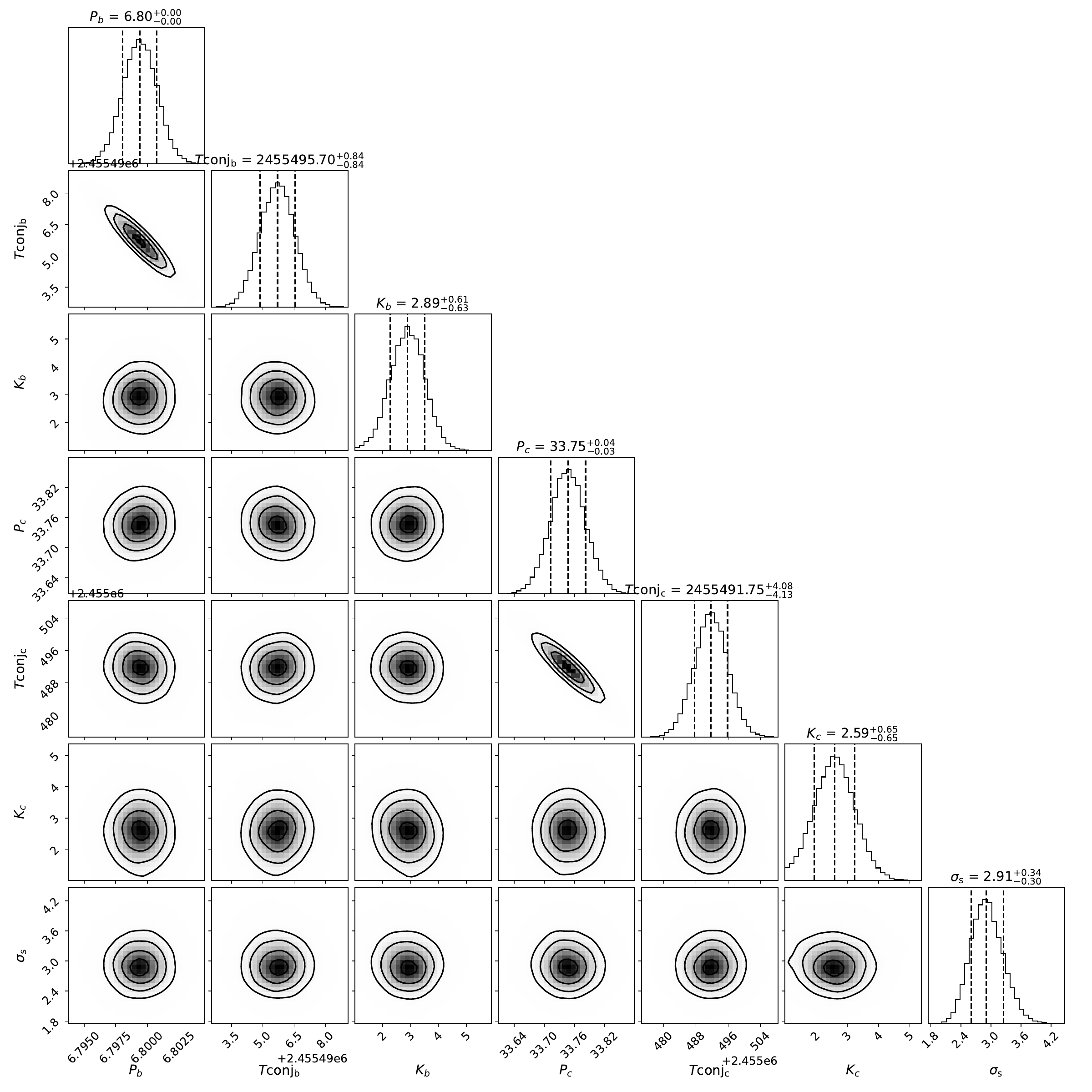}
\caption{Corner plot of derived parameters of the planets around 55 Cnc B assuming the orbits are circular.
}\label{fig:55cncb_rv_corner}
\end{figure*}

Finally, Figure \ref{fig:55cncb_rv_corner} shows the model parameters for the 55 Cnc B planetary system, assuming circular orbits.

\onecolumn
\begin{table}
\caption{Best-fit parameters of the 6-Keplerian signals modelled in the time series of 55 Cnc A. }
\label{tab:planetsA}
    \begin{tabular}{lll}
\hline
    Parameter      & Prior & Posterior \\\hline
  $P_{e}$ (d)      &$\mathcal{N}$(0.73654,0.001) & $0.736546\pm 1e-06$\\
  $T\rm{conj}_{e}$ & $\mathcal{N}$(2455500.25, 0.01)& $2455500.2546\pm 0.005$  \\
  $e_{e}$          &$< 1$ &  $0.01^{+0.02}_{-0.01}$   \\
  $\omega_{e}$ (rad) & & $-0.1^{+1.8}_{-2.3}$   \\
  $K_{e}$ (m s$^{-1}$) & $\mathcal{J}$(0.01, 20)&  $6.55\pm 0.21$ \\
  $a_e$ (au)          &  &  $0.016\pm 0.0003$               \\
  $m_e$sin$i$ (\ME)  &  & $9.35\pm 0.44$              \\\hline
  $P_{b}$ (d)& $\mathcal{N}$(14.6515,0.001)& $14.651552\pm 3.6e-05$  \\
  $T\rm{conj}_{b}$ &$\mathcal{N}$(2455495.5, 0.1) & $2455495.595^{+0.012}_{-0.015}$  \\
  $e_{b}$ & $< 1$& $0.0029^{+0.0027}_{-0.002}$   \\
  $\omega_{b}$ (rad)  & & $0.41^{+0.97}_{-1.0}$   \\
  $K_{b}$  (m s$^{-1}$)& $\mathcal{J}$(0.01, 150)& $71.5\pm 0.22$  \\
  $a_b$ (au) &  &    $0.118\pm0.002$       \\
  $m_b$sin$i$ (\ME)&  &    $276\pm9$             \\\hline
  $P_{c}$ (d)& $\mathcal{N}$(44.39,0.1)& $44.3936\pm 0.0024$ \\
  $T\rm{conj}_{c}$ & $\mathcal{N}$(2455491, 2)& $2455491.23\pm0.34$  \\
  $e_{c}$ &$< 1$ & $0.088\pm0.23$   \\
  $\omega_{c}$ (rad) &  &$0.54\pm 0.24$  \\
  $K_{c}$  (m s$^{-1}$)& $\mathcal{J}$(0.01, 50)& $10.17\pm 0.22$  \\
  $a_c$ (au) &  &    $0.247\pm0.004$       \\
  $m_c$sin$i$ (\ME)&  &   $56.6\pm$2.2              \\\hline
  $P_{f}$ (d)& $\mathcal{N}$(260.59,2)& $260.58\pm0.15$  \\
  $T\rm{conj}_{f}$ & $\mathcal{N}$(2455237, 15) &$2455236.4\pm2.6$  \\
  $e_{f}$ & $< 1$& $0.063^{+0.048}_{-0.043}$  \\
  $\omega_{f}$ (rad) & & $1.75^{+0.79}_{-0.44}$  \\
  $K_{f}$  (m s$^{-1}$)& $\mathcal{J}$(0.01, 50)& $4.83\pm 0.22$  \\
  $a_f$ (au) &  &   $0.802\pm$0.014        \\
  $m_f$sin$i$ (\ME)&  &   $48.5\pm2.8$          \\\hline
  $P_{d}$ (d)&$\mathcal{N}$(4774,100) & $4799.0^{+4.9}_{-15.0}$  \\
  $T\rm{conj}_{d}$ &$\mathcal{N}$(2451558, 100)  &$2451517.8^{+20.0}_{-3.5}$  \\
  $e_{d}$ & $< 1$ &$0.0913\pm0.0067$   \\
  $\omega_{d}$ (rad) & & $-1.57\pm0.04$ \\
  $K_{d}$  (m s$^{-1}$)&$\mathcal{J}$(0.01, 100) & $45.63\pm0.45$   \\
  $a_d$ (au) &  &    $5.60\pm0.10$           \\
  $m_d$sin$i$  (M$_{\rm Jup}$)&  &   $3.82\pm0.13$  \\\hline
  $P_{k6}$ (d)&$\mathcal{N}$(3832,100) & $3827.1^{+9.3}_{-6.4}$ \\
  $T\rm{conj}_{k6}$ &$\mathcal{N}$(2450863, 300)  &$2450841.5^{+12.0}_{-5.8}$ \\
  $e_{k6}$ & $< 1$& $0.35\pm 0.02$  \\
  $\omega_{k6}$ (rad) & & $-1.66\pm0.03$ \\
  $K_{k6}$  (m s$^{-1}$)& $\mathcal{J}$(0.01, 50) &$13.06\pm0.48$  \\
\hline
  $\gamma_{\rm Keck-post}$ (m s$^{-1}$) &-& $-21.8\pm 1.1$  \\
  $\gamma_{\rm Keck-pre}$ (m s$^{-1}$) &-& $-23.07\pm 0.34$\\
  $\gamma_{\rm APF}$ (m s$^{-1}$) &-& $9.56\pm 0.58$   \\
  $\gamma_{\rm Tull}$ (m s$^{-1}$) &-& $-22568.72\pm 0.58$  \\
  $\gamma_{\rm SPIRou}$  (m s$^{-1}$)&-& $27929.0\pm 1.2$  \\
  $\gamma_{\rm Hamilton}$ (m s$^{-1}$) &-& $5.91\pm 0.58$ \\
  $\dot{\gamma}$ (m s$^{-1}$ d$^{-1}$) &-& $0.00232\pm 0.00017$    \\
  $\sigma_{\rm Keck-post}$ (m s$^{-1}$) &$\mathcal{J}$(0.01, 10)& $3.53\pm 0.77$   \\
  $\sigma_{\rm Keck-pre}$ (m s$^{-1}$) &$\mathcal{J}$(0.01, 10)& $3.72\pm 0.24$  \\
  $\sigma_{\rm APF}$ (m s$^{-1}$) &$\mathcal{J}$(0.01, 10)& $3.22\pm 0.28$  \\
  $\sigma_{\rm Tull}$ (m s$^{-1}$) &$\mathcal{J}$(0.01, 10)& $4.14\pm 0.40$ \\
  $\sigma_{\rm SPIRou}$ (m s$^{-1}$) &$\mathcal{J}$(0.01, 10)& $2.94\pm 0.24$  \\
  $\sigma_{\rm Hamilton}$ (m s$^{-1}$) &$\mathcal{J}$(0.01, 10)&  $6.49\pm 0.34$ \\
  \hline
  \end{tabular}
  \tablefoot{{$k6$ stands for the 6$^{th}$ Keplerian signal. $\mathcal{J}$ stands for a Jeffreys prior and $\mathcal{N}$ for a normal prior.}}
\end{table}

\end{appendix}

\end{document}